\documentclass[a4paper, 10pt]{article}

\usepackage[style=apa, 
backend=biber, 
uniquename=false,
url=true, 
doi=true]{biblatex}
\usepackage{hyperref}
\usepackage[dvipsnames]{xcolor}
\usepackage{colortbl}
\hypersetup{
plainpages=false,%
bookmarksopen=true,%
bookmarksnumbered=false,%
bookmarksdepth=5,
breaklinks=true,
colorlinks=true,
citecolor=BlueViolet,
linkcolor=BlueViolet,
urlcolor=blue,
}
\usepackage{citeauthor}

\usepackage{mathrsfs} 

\usepackage[T1]{fontenc}
\usepackage[utf8]{inputenc}

\usepackage{times} 

\usepackage{bm}
\usepackage{amsmath}
\usepackage{amssymb}

\usepackage{SpringerPsychometrika}

\usepackage{fancyhdr}

\fancypagestyle{firstpage}{
\fancyhf{}
\rhead{\textit{arXiv Preprint}}
\lhead{}
\chead{}
\cfoot{}
\lfoot{Published at \textit{Behaviormetrika} in 2022}

}

\fancypagestyle{fund_footer}{%
\fancyhf{}

\fancyfoot[L]{\small}
}

\usepackage{setspace}

\usepackage[noblocks, affil-it]{authblk}

\usepackage{abstract}
\renewenvironment{abstract}{\section*{\centering\abstractname}}{}
\renewcommand{\abstractname}{}

\usepackage{graphicx}
\newcommand{\orcid}[1]{\href{#1}{\includegraphics[width=10pt]{orcid.png}}}

\usepackage{graphicx}

\usepackage{booktabs}
\usepackage{multicol}
\usepackage{multirow}
\usepackage{dcolumn}
\usepackage{dcolumn}
\usepackage{lscape}

\newcolumntype{L}{>{$}l<{$}} 
\newcolumntype{C}{>{$}c<{$}} 
\newcolumntype{R}{>{$}r<{$}} 

\usepackage{float}

\usepackage{threeparttable}
\usepackage{adjustbox}
\usepackage{algorithm}
\usepackage{algorithmic}

\begin{document}

\title{\vspace{-1.5\bigskipamount} \normalsize \uppercase{Variational Bayesian Inference for a Polytomous-Attribute Saturated Diagnostic Classification Model with Parallel Computing}}
\author[ ]{\fontsize{11pt}{11pt}{\scshape Motonori Oka}}
\author[ ]{\fontsize{11pt}{11pt}{\scshape Shun Saso}}
\author[ ]{\fontsize{11pt}{11pt}{\scshape Kensuke Okada}}
\affil[ ]{\small{GRADUATE SCHOOL OF EDUCATION, THE UNIVERSITY OF TOKYO}}

\date{}
\maketitle

\thispagestyle{firstpage}

\vspace{-4\bigskipamount}
\begin{abstract}
\begin{center}\footnotesize
\begin{minipage}{\dimexpr\paperwidth-79mm}
\setlength{\parindent}{12pt}
As a statistical tool to assist formative assessments in educational settings, diagnostic classification models (DCMs) have been increasingly used to provide diagnostic information regarding examinees' attributes. DCMs often adopt a dichotomous division such as the mastery and non-mastery of attributes to express the mastery states of attributes. However, many practical settings involve different levels of mastery states rather than a simple dichotomy in a single attribute. Although this practical demand can be addressed by polytomous-attribute DCMs, their computational cost in a Markov chain Monte Carlo estimation impedes their large-scale application due to the larger number of polytomous-attribute mastery patterns than that of binary-attribute ones. This study considers a scalable Bayesian estimation method for polytomous-attribute DCMs and developed a variational Bayesian (VB) algorithm for a polytomous-attribute saturated DCM---a generalization of polytomous-attribute DCMs---by building on the existing literature on polytomous-attribute DCMs and VB for binary-attribute DCMs. Furthermore, we proposed the configuration of parallel computing for the proposed VB algorithm to achieve better computational efficiency. Monte Carlo simulations revealed that our method exhibited the high performance in parameter recovery under a wide range of conditions. An empirical example is used to demonstrate the utility of our method. \\

\end{minipage}
\end{center}
\end{abstract}

\vspace{\bigskipamount}

The recent demand for formative assessments has motivated the development of psychometric models providing diagnostic information on students' latent traits. A class of these models is termed \emph{diagnostic classification models} (DCMs). It postulates that latent traits generally consist of multidimensional discrete skills called ``attributes,'' and students belong to one of the mastery profiles for these attributes, each of which presents a combination of the mastery and non-mastery of attributes \parencite{rupp_diagnostic_2010}. Based on this assumption, DCMs aim to estimate the mastery probabilities of attributes for every individual and classify them into attribute mastery profiles. Thanks to this information, practitioners can design classroom activities that emphasize students' educational needs.

Although most of the DCMs assign binary attributes in their parameterization, a few DCMs with polytomous attributes have been developed to perform more finer-grained diagnoses \parencite[for example,][]{tzur_karelitz_ordered_2004,templin_generalized_2004,vondavier_general_2008,templin_measuring_2013,chen_general_2013,von_davier_mixture_2007}. The seminal contribution to polytomous-attribute DCMs is Karelitz's (\citeyear{tzur_karelitz_ordered_2004}) work, which introduced the ordered-category attribute coding framework to treat different attribute mastery levels in an attribute. In contrast to binary attributes that only consider a dichotomous division between their mastery and non-mastery, this framework broadens the dichotomous treatment and addresses the qualitative ordering of cognitive complexity within an attribute. For instance, \textcite{tjoe_identification_2014} incorporated polytomous attributes as a part of the specified attributes for proportional reasoning, where ``constructing ratios'' and ``constructing proportions'' are assumed to present different mastery levels. Both attributes fall within the scope of a ratio. However, constructing proportions is more difficult to master than constructing ratios because calculating proportions requires the understanding of ratios. The inclusion of such attributes forms a new class of attribute mastery profiles untouched by binary attributes and helps practitioners measure students' levels of mastery over attributes with more refined granularity. 

This study considers the polytomous-attribute extension of an ordinary binary-attribute saturated DCM. A saturated DCM is a generalization of DCMs in the sense that all the main and interaction effects of the attributes are allowed, and each possible attribute mastery profile can have a unique correct-response probability. Examples of such binary-attribute models include the general diagnostic model \parencite[GDM:][]{vondavier_general_2008}, log-linear cognitive diagnostic model \parencite[LCDM:][]{henson_defining_2009}, and generalized deterministic input, noisy ``and'' gate (G-DINA) model \parencite{delatorre_generalized_2011}. For the generalized model of polytomous-attribute DCMs, \textcite{chen_general_2013} developed the polytomous G-DINA (pG-DINA) model by introducing a method to reduce polytomous-attribute mastery profiles to item-specific binary profiles pertaining to the unique correct-response probabilities for an item. This enables estimating the parameters of the pG-DINA model in the same manner as the G-DINA model.

Regardless of polytomy in an attribute, a class of saturated DCMs holds substantial utility. First, any sub-models in DCMs such as the deterministic input, noisy ``and'' gate \parencite[DINA;][]{junker_cognitive_2001} model,
deterministic input, noisy ``or'' gate \parencite[DINO;][]{templin_measurement_2006} model, reduced reparameterized unified model \parencite[RRUM;][]{hartz_fusion_2008}, and compensatory RUM \parencite[CRUM;][]{rupp_diagnostic_2010} can be expressed within a saturated DCM by imposing appropriate restrictions on its parameters. For example, a saturated DCM with only the intercept and highest-order interaction terms of required attributes allowed is effectively the DINA model, which assumes that examinees must master all the necessary attributes to answer an item correctly. This flexibility in modeling different types of item-responding processes brings about the second advantage. Since a saturated DCM nests its sub-models, we can infer the nature of the item-responding process behind an item by inspecting its item parameter estimates. As is the case with the DINA model, if the intercept and highest-order interaction terms of a saturated DCM are estimated to be high, and other terms are estimated to be close to zero, then the item-responding process behind a corresponding item is likely to be the DINA model. As such, the item parameters in a saturated DCM richly portray the item-responding processes behind the items, informing practitioners quantitatively about how examinees respond to given items.

Despite the merits of a saturated DCM mentioned above, the computational speed for its Bayesian estimation can be painstakingly slow at large-scale settings where the numbers of examinees and attributes are sizable. This is because Markov chain Monte Carlo (MCMC)---a commonly used method for Bayesian estimation---is generally not scalable to such large-scale settings owing to its stochastic search for the parameter space of the targeted posterior \parencite{gelman_bayesian_2013}. To reduce this computational cost, a variational Bayesian (VB) inference method is often employed as an alternative to an MCMC estimation. VB inference is a deterministic and fast approach to approximate posterior distributions. This deterministic nature in an estimation procedure enables a scalable Bayesian estimation, and many applied researchers have utilized it for their Bayesian modeling to capitalize on its high scalability \parencite{blei_variational_2017}. In DCM literature, several VB inference algorithms have been developed in recent years \parencite{yamaguchi_variational_2020,yamaguchi_variational_2021,yamaguchi_vbmultiplechoice_2020}. In particular, \textcite{yamaguchi_variational_2021} developed a VB algorithm for a binary-attribute saturated DCM by introducing a G-matrix to reformulate it as a Bernoulli mixture model so that the priors for its model parameters become conditionally conjugate; this conditional conjugacy simplifies the derivation of a coordinate ascent mean-field VB inference \parencite{wang_variational_2013}. Additionally, \textcite{yamaguchi_variational_2021} confirmed the sound accuracy and fast computation of their VB algorithm and showed the superior computational efficiency of VB estimation over maximum likelihood estimation based on an expectation-maximization (EM) algorithm under the condition that the response data are obtained consecutively, such as in computerized adaptive testing.

By contrast, none of these studies have worked on the problem of the scalability for polytomous-attribute DCMs despite the fact that these models would present a more severe computational problem in application settings. This particular severity in their MCMC computation results from the nature of polytomous attributes, in which the increase of mastery levels generally leads to expanding the parameter space of attribute mastery profiles from $2^K$ to $M^K$. Here, $K$ and $M$ denote the number of attributes and mastery levels, respectively. This expansion causes more intensive MCMC computation compared to binary-attribute DCMs. Therefore, it is desirable to develop a scalable Bayesian estimation algorithm for a polytomous-attribute saturated DCM. Accordingly, in this study, we develop a VB algorithm for a polytomous-attribute saturated DCM, which builds on the prior work by \textcite{chen_general_2013} and \textcite{yamaguchi_variational_2021}, and propose its parallelized algorithm based on the configuration of the parallel-E parallel-M algorithm for generalized latent variable models \parencite{von_davier_high-performance_2016}. Furthermore, we conduct simulation and empirical studies to assess the utility of our algorithm.

The remainder of this paper is structured as follows. In Section 2, we first introduce the two foundational models of \textcite{chen_general_2013} and \textcite{yamaguchi_variational_2021} and then explain the polytomous-attribute G-matrix to derive a VB algorithm for a polytomous-attribute saturated DCM and the configuration of its parallel computing. Subsequently, our simulation and empirical studies are presented in Sections 3 and 4. Lastly, we discuss the limitations and future directions of this study.

\section{Methods}
\subsection{pG-DINA Model}\label{sec:2.1}
Let $i\;(1,\ldots,N)$, $j\;(1,\ldots,J)$, and $k\;(1,\ldots,K)$ express respondents, items, and attributes, respectively. For ease of notation, we assume that all the attributes have the same mastery level $M_k=M$. The number of attribute mastery profiles for polytomous-attribute DCMs is $M^K$, each of which is denoted as $l\;(1,\ldots,L=M^K)$. An implementation of DCMs requires a $J\times K$ Q-matrix that specifies the relationships between the items and attributes. An entry $q_{jk}$ of a Q-matrix can take values from $0$ to $M-1$. If the value of $q_{jk}$ is $m\;(0,\ldots,M-1)$, more than the $m$-th level of mastery in attribute $k$ is necessary for a correct response to item $j$. A set of relevant attributes for item $j$ is indicated by $\bm{q}_{j}=(q_{j1}, \ldots, q_{jK})^\mathsf{T}$, which is the $j$-th vector of a Q-matrix \textcolor{black}{$\mathbf{Q}=(\bm{q}_1, \ldots, \bm{q}_J)^\mathsf{T}$.} In addition, let $\bm{\alpha}_l=(\alpha_{l1}, \ldots, \alpha_{lK})^\mathsf{T}$ be the $l$-th vector of an $L\times K$ attribute mastery profile matrix \textcolor{black}{$\mathbf{A}_{\mathrm{profile}}=(\bm{\alpha}_1, \ldots, \bm{\alpha}_L)^\mathsf{T}$} that comprises all the possible attribute mastery profiles. An entry of $\bm{\alpha}_l$ can also take values from 0 to $M-1$. The superscript $\mathsf{T}$ represents the transpose.

Moreover, following the notations adopted in \textcite{chen_general_2013} and \textcite{delatorre_generalized_2011}, we use $K_j^* = \sum_{k=1}^K I(q_{jk}>0)$ to denote the number of relevant attributes for item $j$. $I(\cdot)$ is an indicator function that takes the value of 1 when a given condition is satisfied. Using the notation of $K_j^*$, we can reduce each vector of an attribute mastery profile matrix \textcolor{black}{$\mathbf{A}_{\mathrm{profile}}$} to the reduced vector $\bm{\alpha}_{jl}^*=(\alpha_{jl1}^*, \ldots, \alpha_{jlK_j^*}^*)^\mathsf{T}$ for item $j$, where $l=1, \ldots, M^{K_j^*}$ and each entry of $\bm{\alpha}_{jl}^*$ corresponds to the attributes satisfying $q_{jk}>0$ on a vector $\bm{q}_j$. Besides, \textcite{chen_general_2013} modified the reduced vector to the $\textit{collapsed attribute vector}$ $\bm{\alpha}_{jl}^{**}=(\alpha_{jl1}^{**}, \ldots, \alpha_{jlK_j^*}^{**})^\mathsf{T}$ for further simplification, where each element of the reduced attribute vector $\bm{\alpha}_{lj}^*$ is collapsed into a binary element:
\begin{align}
\alpha^{**}_{jlk} = 
\begin{cases}
0 \;\text{if}\; \alpha^{*}_{jlk} < q_{jk} & \\
 1 \;\text{otherwise} &
\end{cases}.
\end{align}
The example of this transformation based on the one shown in \textcite{chen_general_2013} is illustrated in Table \ref{tab:colappsedvec}. Consider the case of $K=3$, $M=3$, and the item with $\bm{q}_j=(2, 1, 0)^\mathsf{T}$. Since $K^*_j$ for this $\bm{q}$ vector is 2, the number of unique reduced attribute vectors $\bm{\alpha}_{lj}^*$ is $M^{K^*_j}=3^2=9$. Those vectors can be further collapsed into $L_j^*=2^{K^*_j}=2^2=4$ unique collapsed attribute vectors $\bm{\alpha}_{lj}^{**}$. These collapsed vectors are essentially the patterns that can be distinguished by item $j$.

\begin{table}[!htbp]
\centering
\caption{The example of collapsed attribute vectors when $\bm{q}_j=(2,1,0)$}
 \begin{tabular}{ccc}
 \toprule
 Original $\bm{\alpha}_l$ & Reduced $\bm{\alpha}^*_{jl}$ & Collapsed $\bm{\alpha}_{jl}^{**}$ \\
 \midrule
 $(0,0,0), (0,0,1), (0,0,2)$ & $(0,0)$ & \multirow{2}{*}{$(0,0)$} \\
 $(1,0,0), (1,0,1), (1,0,2)$ & $(1,0)$ &\\
 & & \\
 $(2,0,0), (2,0,1), (2,0,2)$ & $(2,0)$ & $(1,0)$ \\
 & & \\
 $(0,1,0), (0,1,1), (0,1,2)$ & $(0,1)$ & \multirow{4}{*}{$(0,1)$} \\
 $(1,1,0), (1,1,1), (1,1,2)$ & $(1,1)$ &\\
 $(0,2,0), (0,2,1), (0,2,2)$ & $(0,2)$ &\\ 
 $(1,2,0), (1,2,1), (1,2,2)$ & $(1,2)$ &\\
 & & \\
 $(2,1,0), (2,1,1), (2,1,2)$ & $(2,1)$ & \multirow{2}{*}{$(1,1)$} \\
 $(2,2,0), (2,2,1), (2,2,2)$ & $(2,2)$ &\\
 \bottomrule
 \end{tabular}%
\label{tab:colappsedvec}%
\end{table}%

Accordingly, the number of unique correct-response probabilities for this item becomes 4, and each of them is expressed by the item response function of the pG-DINA model:
\begin{align}
&P(x_{ij}=1 \vert \bm{\alpha}^{**}_{jl}) \nonumber\\
&\quad = \delta_{j0} + \sum_{k=1}^{K^*_j}\delta_{jk}\alpha_{jlk}^{**} + \sum_{k'>k}^{K^*_j}\sum_{k=1}^{K^*_j}\delta_{jkk'}\alpha_{jlk}^{**}\alpha_{jlk'}^{**} + \cdots + \delta_{j1,\ldots,K^*_j}\prod_{k=1}^{K^*_j}\alpha_{jlk}^{**},
\end{align}
where $l=1, \ldots, L_j^*(=2^{K^*_j})$. Here, $\delta_{j0}$ and $\delta_{jk}$ denote the intercept and main-effect terms of the relevant attributes, respectively. The terms after the second one denote all the combinations of the interaction effects of the relevant attributes. With the simplified formulation of attribute mastery profiles elaborated in Table \ref{tab:colappsedvec}, the item response function of the pG-DINA model is formulated to be the same as the G-DINA model, except for the fact that the attribute mastery profiles embedded in the item response function of the pG-DINA model are the collapsed attribute vectors, whereas those of the G-DINA model are the binary reduced attribute vectors \parencite{chen_general_2013}. Because of such equivalence in their item response functions, the estimation procedure of the G-DINA model can be applied directly to that of the pG-DINA model.

\subsection{Binary-Attribute Saturated DCM Using G-Matrices}\label{sec:2.2}
Since each item possesses its own correct-response probabilities associated with item-specific attribute mastery profiles, it can be construed that a set of binary responses $\bm{x}_j$ for item $j$ is generated from heterogeneous populations with different Bernoulli distribution functions, where each item-specific attribute mastery profile has a unique Bernoulli probability. Based on this insight, \textcite{yamaguchi_variational_2021} leveraged the idea of mixture modeling for heterogeneous populations and reformulated a binary-attribute saturated DCM as a Bernoulli mixture model by introducing a \textit{latent indicator \textcolor{black}{vector}} $\bm{z}_i$ and a \textit{G-matrix} $\mathbf{G}_j$.

In this section, for improved clarity of notations between binary-attribute and polytomous-attribute DCMs, we use $H=2^K$ to denote the number of attribute mastery profiles for binary-attribute DCMs, although we use $L=M^K$ to denote that for polytomous-attribute DCMs in Section \ref{sec:2.1}. \textcolor{black}{In addition, let $h^*\; (1,\ldots,H^*_j)$ and $l^*\; (1,\ldots,L^*_j)$ represent the indices for the item-specific number of reduced attribute vectors in binary-attribute DCMs and that of collapsed attribute vectors in polytomous-attribute DCMs, respectively.} A latent indicator \textcolor{black}{vector} $\bm{z}_i =(z_{i1}, \ldots, z_{iH})^\mathsf{T}$ specifies the attribute mastery pattern to which examinee $i$ belongs and is a vector with $H=2^K$ elements, where the $h$-th element takes the value of 1 when examinee $i$ belongs to class $h$ so that the values of $z_{ih}$ satisfy $z_{ih}=1$ and $\sum_{h=1}^H z_{ih}=1$. A G-matrix $\mathbf{G}_j$, which is the $H_j^*(= 2^{K^*_j}) \times H$ matrix, is introduced in \textcite{yamaguchi_variational_2021} to reduce a latent indicator \textcolor{black}{vector} $\bm{z}_i$ to an item-specific latent indicator \textcolor{black}{vector} $\bm{z}_{ji}=(z_{ji1}, \ldots, z_{jiH_j^*})^\mathsf{T}$ with $H_j^*$ elements, indicating to which item-specific attribute mastery pattern examinee $i$ belongs. The example of a G-matrix $\mathbf{G}_j$ based on \textcite{yamaguchi_variational_2021} is illustrated in Table \ref{tab:G-matrix}. 
\begin{table}[!htbp]
\centering
\caption{The example of a G-matrix when $\bm{q}_j=(1,1,0)$}
\resizebox{\linewidth}{!}{
 \begin{tabular}{rccccccccccc}
 \toprule
&&&& \multicolumn{8}{c}{Attribute mastery patterns} \\
\cmidrule{5-12}&&&& \multicolumn{1}{c}{1} & \multicolumn{1}{c}{2} & \multicolumn{1}{c}{3} & \multicolumn{1}{c}{4} & \multicolumn{1}{c}{5} & \multicolumn{1}{c}{6} & \multicolumn{1}{c}{7} & 8 \\
 \midrule
&&& $\alpha_1$ & \multicolumn{1}{c}{0} & \multicolumn{1}{c}{1} & \multicolumn{1}{c}{0} & \multicolumn{1}{c}{0} & \multicolumn{1}{c}{1} & \multicolumn{1}{c}{1} & \multicolumn{1}{c}{0} & 1 \\
&&& $\alpha_2$ & \multicolumn{1}{c}{0} & \multicolumn{1}{c}{0} & \multicolumn{1}{c}{1} & \multicolumn{1}{c}{0} & \multicolumn{1}{c}{1} & \multicolumn{1}{c}{0} & \multicolumn{1}{c}{1} & 1 \\
&&& $\alpha_3$ & \multicolumn{1}{c}{0} & \multicolumn{1}{c}{0} & \multicolumn{1}{c}{0} & \multicolumn{1}{c}{1} & \multicolumn{1}{c}{0} & \multicolumn{1}{c}{1} & \multicolumn{1}{c}{1} & 1 \\
 \midrule
 Item-specific attribute mastery patterns & $\alpha_1$ & $\alpha_2$ & $\alpha_3$ & \multicolumn{8}{c}{G-matrix} \\
 \midrule
 1& 0& 0& *& 1& 0& 0& 1& 0& 0& 0& 0 \\
 2& 1& 0& *& 0& 1& 0& 0& 0& 1& 0& 0 \\
 3& 0& 1& *& 0& 0& 1& 0& 0& 0& 1& 0 \\
 4& 1& 1& *& 0& 0& 0& 0& 1& 0& 0& 1 \\
 \bottomrule
 \end{tabular}%
 }
\label{tab:G-matrix}%
\end{table}%

Consider the case of $K=3$ and the item with $\bm{q}_j=(1, 1, 0)^\mathsf{T}$. Since the number of \textcolor{black}{attributes measured by this item} is $K^*_j = 2$, the number of reduced attribute mastery patterns becomes $H^*_j=2^{K^*_j} = 4$. Thus, $\mathbf{G}_j$ becomes a $4 \times 8$ matrix. With this matrix, an item-specific latent indicator vector $\bm{z}_{ji}$ is computed by multiplying $\mathbf{G}_j$ by $\bm{z}_i$ such that $\bm{z}_{ji}=\mathbf{G}_j\bm{z}_i$. Elements of $\bm{z}_{ji}$ are $z_{jih^*}=\sum_{h=1}^{H}g_{jh^*h}z_{jih}$, where $h^*=1, \ldots, H^*_j(=2^{K_j^*})$ and $h=1,\ldots,H(=2^K)$. These elements take the value of 1 when examinee $i$ belongs to the $h^*$-th item-specific attribute mastery pattern and satisfy $\sum_{h^*=1}^{H_j^*}z_{jih^*}=1$. For instance, under the specification in Table \ref{tab:G-matrix} and $\bm{z}_i =(0,0,0,0,0,0,0,1)^\mathsf{T}$, $\bm{z}_{ji}=\mathbf{G}_j\bm{z}_i$ is computed as $\bm{z}_{ji}=(0,0,0,1)^\mathsf{T}$, where $\mathbf{G}_j$ correctly converts $\bm{z}_i$ to an item-specific latent indicator vector $\bm{z}_{ji}$ that has the value of 1 on its fourth element.
Based on $\mathbf{G}_j$ and $\bm{z}_i$, \textcite{yamaguchi_variational_2021} formulated the item response function of the binary-attribute saturated DCM with G-matrices as follows:
\begin{align}
P(x_{ij}=1 \vert \bm{z}_i, \bm{\theta}_j, \mathbf{G}_j, \bm{q}_j) =\prod_{h^*=1}^{H^*_j}\theta_{jh^*}^{z_{jih^*}},
\end{align}
where $\theta_{jh^*}$ denotes the correct-response probability of item $j$ for an examinee with the $h^*$-th item-specific attribute mastery pattern.
In addition, under the assumption of local independence given a latent indicator \textcolor{black}{vector} $\bm{z}_i$, \textcite{yamaguchi_variational_2021} provided the likelihood function of the binary-attribute saturated DCM using G-matrices:
\begin{align}
P(\mathbf{X} \vert \mathbf{Z}, \mathbf{\Theta}, \mathbf{G}, \mathbf{Q}) &= \prod_{i=1}^{N}\prod_{j=1}^{J}\prod_{h^*=1}^{H^*_j}P(x_{ij} \vert \bm{z}_i, \bm{\theta}_j, \mathbf{G}_j, \bm{q}_j) \nonumber\\
&= \prod_{i=1}^{N}\prod_{j=1}^{J}\prod_{h^*=1}^{H^*_j} \left\{\theta_{jh^*}^{x_{ij}}(1-\theta_{jh^*})^{1-x_{ij}}\right\}^{z_{jih^*}}.
\end{align}
Since the above likelihood function is in a Bernoulli mixture formulation, we can apply the well-known procedure of VB inference for Bernoulli mixture models.

\subsection{Formulate a G-Matrix for a Polytomous-Attribute Saturated DCM}\label{sec:2.3}

\textcolor{black}{Similar to a G-matrix for the binary-attribute saturated DCM in Section \ref{sec:2.2}, a polytomous-attribute G-matrix is obtained by introducing collapsed attribute vectors into the construction of G-matrices. An example of a polytomous-attribute G-matrix based on collapsed attribute vectors is shown in Table \ref{tab:polyG-matrix}. In the same manner as a binary-attribute G-matrix, a polytomous-attribute G-matrix $\mathbf{G}_j^{\mathrm{poly}}$ converts a latent indicator vector for polytomous attributes $\bm{z}_i =(z_{i1}, \ldots, z_{iL})^\mathsf{T}$ to an item-specific indicator vector based on collapsed attribute vectors $\bm{z}_{ji}=(z_{ji1}, \ldots, z_{jiL_j^*})^\mathsf{T}$.}

\begin{table}[!htbp]
\color{black}
\centering
\caption{The example of a polytomous-attribute G-matrix \textcolor{black}{based on collapsed attribute vectors} when $K=2$, $M=3$, and $\bm{q}_j=(2,0)$}
\begin{tabular}{rccccccccccc}
\toprule
& & & \multicolumn{9}{c}{Attribute mastery patterns} \\
\cmidrule{4-12}& & & 1 & 2 & 3 & 4 & 5 & 6 & 7 & 8 & 9 \\
\midrule
& & $\alpha_1$& 0 & 1 & 0 & 1 & 2 & 0 & 1 & 2 & 2 \\
& & $\alpha_2$& 0 & 0 & 1 & 1 & 0 & 2 & 2 & 1 & 2 \\
\midrule
Collapsed attribute mastery patterns & $\alpha_1$& $\alpha_2$& \multicolumn{9}{c}{Polytomous-attribute G-matrix} \\
\midrule
1 & 0 & * & 1 & 1 & 1 & 1 & 0 & 1 & 1 & 0 & 0 \\
2 & 1 & * & 0 & 0 & 0 & 0 & 1 & 0 & 0 & 1 & 1 \\
\bottomrule
\end{tabular}%
\label{tab:polyG-matrix}%
\end{table}%

Based on $\mathbf{G}_j^{\mathrm{poly}}$ and $\bm{z}_i$, the item response function of the polytomous-attribute saturated DCM is defined as
\begin{align}
P(x_{ij}=1 \vert \bm{z}_i, \bm{\theta}_j, \mathbf{G}_j^{\mathrm{poly}}, \bm{q}_j) =\prod_{l^*=1}^{L^*_j}\theta_{jl^*}^{z_{jil^*}}.
\end{align}
Its likelihood function is also defined as
\begin{align}
P(\mathbf{X} \vert \mathbf{Z}, \mathbf{\Theta}, \mathbf{G}^{\mathrm{poly}}, \mathbf{Q}) &= \prod_{i=1}^{N}\prod_{j=1}^{J}\prod_{l^*=1}^{L^*_j}P(x_{ij} \vert \bm{z}_i, \bm{\theta}_j, \mathbf{G}_j^{\mathrm{poly}}, \bm{q}_j) \nonumber\\
&= \prod_{i=1}^{N}\prod_{j=1}^{J}\prod_{l^*=1}^{L^*_j} \left\{\theta_{jl^*}^{x_{ij}}(1-\theta_{jl^*})^{1-x_{ij}}\right\}^{z_{jil^*}}.
\end{align}
As \textcite{yamaguchi_variational_2021} noted for the binary-attribute saturated DCM using G-matrices, the intercept, main, and interaction effects of the attributes in the pG-DINA model can also be obtained by transforming the estimated $\bm{\theta}_j$ to $\bm{\delta}_j$ using the least-square estimation elaborated in \textcite{delatorre_generalized_2011}.

{\color{black}
\subsubsection{Extending a Polytomous-Attribute G-Matrix with Collapsed Attribute Vectors to that with Reduced Attribute Vectors}
The utility of a polytomous-attribute G-matrix is not limited only to modeling correct-response probabilities based on collapsed attribute vectors. A G-matrix based on reduced attribute vectors can also be formulated, allowing us to flexibly capture the subtle differences in correct-response probabilities at different levels of mastery states. An example of a polytomous-attribute G-matrix using reduced attribute vectors with $K=2$, $M=3$, and $\bm{q}_j=(2,0)$ is presented in Table \ref{tab:reducedpolyG-matrix}.

\begin{table}[!htbp]
\color{black}
\centering
\caption{The example of a polytomous-attribute G-matrix based on reduced attribute vectors when $K=2$, $M=3$, and $\bm{q}_j=(2,0)$}
\begin{tabular}{rccccccccccc}
\toprule
& & & \multicolumn{9}{c}{Attribute mastery patterns} \\
\cmidrule{4-12}& & & 1 & 2 & 3 & 4 & 5 & 6 & 7 & 8 & 9 \\
\midrule
& & $\alpha_1$& 0 & 1 & 0 & 1 & 2 & 0 & 1 & 2 & 2 \\
& & $\alpha_2$& 0 & 0 & 1 & 1 & 0 & 2 & 2 & 1 & 2 \\
\midrule
Reduced attribute mastery patterns & $\alpha_1$& $\alpha_2$& \multicolumn{9}{c}{Polytomous-attribute G-matrix} \\
\midrule
1 & 0 & * & 1 & 0 & 1 & 0 & 0 & 1 & 0 & 0 & 0 \\
2 & 1 & * & 0 & 1 & 0 & 1 & 0 & 0 & 1 & 0 & 0 \\
3 & 2 & * & 0 & 0 & 0 & 0 & 1 & 0 & 0 & 1 & 1 \\
\bottomrule
\end{tabular}%
\label{tab:reducedpolyG-matrix}%
\end{table}%

Although the G-matrix based on collapsed attribute vectors in Table \ref{tab:polyG-matrix} assigns the same correct-response probability to respondents with $\alpha_{1}=0$ and those with $\alpha_{1}=1$, the G-matrix using reduced attribute vectors in Table \ref{tab:reducedpolyG-matrix} allows such respondents to have different correct-response probabilities. This G-matrix with the more relaxed assumption of item-responding processes enables a finer-grained inspection of how correct-response probabilities change with different levels of attribute mastery states.}

\subsection{Bayesian Formulation}
\textcolor{black}{In this section, we provide the Bayesian formulation for the polytomous-attribute saturated DCM with its G-matrix using collapsed attribute vectors. Its formulation for the saturated DCM with a polytomous-attribute G-matrix using reduced attribute vectors is the same as the one based on collapsed attribute vectors, except for the number of item-specific attribute mastery patterns with different correct-response probabilities for the items.}

Because the difference in the formulation between the binary- and polytomous-attribute saturated DCMs lies only in the specification of a G-matrix, we follow the Bayesian formulation of the binary-attribute saturated DCM in \textcite{yamaguchi_variational_2021}.

We first write the distribution of $\mathbf{Z}$ as a categorical distribution with the mixing proportions $\bm{\pi}=(\pi_1, \ldots, \pi_L)^{\mathsf{T}}$, which is given as
\begin{align}
P(\mathbf{Z} \vert \bm{\pi}) = \prod_{i=1}^N\prod_{l=1}^L \pi_{l}^{z_{il}}.
\end{align}
The prior over $\bm{\pi}$ is selected to be a Dirichlet distribution with the parameter $\bm{\delta}^0=(\delta^0_1, \ldots, \delta^0_L)^{\mathsf{T}}$:
\begin{align}
P(\bm{\pi} \vert \bm{\delta}^0) = \prod_{l=1}^L \pi_{l}^{\delta_{l}^0-1}.
\end{align}
We choose a Beta distribution for the prior over the correct-response probability parameter $\theta_{jl^*}$:
\begin{align}
P(\theta_{jl^*} \vert a_{jl^*}^{0}, b_{jl^*}^{0}) \propto \theta_{jl^*}^{a_{jl^*}^{0}-1}(1-\theta_{jl^*})^{b_{jl^*}^{0} -1},
\end{align}
where $(a_{jl^*}^{0}, b_{jl^*}^{0})$ are the hyperparameters for $\theta_{jl^*}$. In addition, the assumption of conditional independence on the correct-response probability parameters enables the joint probability of $\theta_{jl^*}$ to be formulated as follows:
\begin{align}
P(\mathbf{\Theta}\vert \mathbf{A}^{0}, \mathbf{B}^{0}) \propto \prod^{J}_{j=1}\prod^{L^*_j}_{l^*=1}\theta_{jl^*}^{a_{jl^*}^{0}-1}(1-\theta_{jl^*})^{b_{jl^*}^{0} -1}.
\end{align}

Based on the likelihood and priors, the joint posterior can be obtained as 
\begin{align}
&P(\mathbf{Z}, \mathbf{\Theta}, \bm{\pi} \vert \mathbf{X}, \mathbf{G}^{\mathrm{poly}}, \mathbf{Q}, \bm{\delta}^0, \mathbf{A}^{0}, \mathbf{B}^{0}) \nonumber\\
&\quad \propto P(\mathbf{X} \vert \mathbf{Z},\mathbf{\Theta},\mathbf{G}^{\mathrm{poly}}, \mathbf{Q})P(\mathbf{Z} \vert \bm{\pi})P(\bm{\pi} \vert \bm{\delta}^0)P(\mathbf{\Theta} \vert \mathbf{A}^{0}, \mathbf{B}^{0}).
\end{align}

\subsection{Variational Bayesian Inference}
As \textcite{blitzstein_introduction_2019} said ``conditioning is the soul of statistics” (p. 46), the quintessence of Bayesian statistics is to update beliefs in current reasoning toward phenomena of interest based on observations. The mathematical formulation of this update is expressed as the Bayes rule in Equation (12), where the posterior distributions capture the uncertainty of parameters:
\begin{eqnarray}
P(\mathbf{\Psi}\vert \mathbf{X}) = \frac{P(\mathbf{X}\vert \mathbf{\Psi})P(\mathbf{\Psi})}{\int P(\mathbf{X}\vert \mathbf{\Psi})P(\mathbf{\Psi})d\mathbf{\Psi}}.
\end{eqnarray}
Although the Bayes rule offers the base of principled statistical inference, computing posterior distributions requires the marginalization term in the denominator. The closed-form solution for this term is usually unavailable in practice, and its numerical integration incurs a prohibitive computational cost \parencite{bishop_pattern_2006}.

VB inference transforms intractable posterior computations into an optimization problem in which the objective is to search for the parametric form of distributions, often referred to as ``variational distributions,'' that produces the best approximation to posteriors \parencite{galdo_variational_2020}. In many cases, Kullback--Leibler (KL) divergence is selected as a metric to measure the discrepancy between the variational distribution $q(\mathbf{\Psi})$ and posterior distribution $P(\mathbf{\Psi}\vert \mathbf{X})$. KL divergence emerges from the well-known decomposition of the log marginal likelihood:
\begin{align}
\log P(\mathbf{X}) &= \int q(\mathbf{\Psi})\log\frac{P(\mathbf{X},\mathbf{\Psi})}{q(\mathbf{\Psi})}d\mathbf{\Psi} - \int q(\mathbf{\Psi})\log\frac{P(\mathbf{\Psi} \vert \mathbf{X})}{q(\mathbf{\Psi})}d\mathbf{\Psi} \nonumber\\
&= L(q) + \mathrm{KL}[ q(\mathbf{\Psi})\parallel P(\mathbf{\Psi}\vert \mathbf{X})] \nonumber\\
&\geq L(q).
\end{align}
Here, $L(q)$ is the lower bound of the log marginal likelihood (variational lower bound; VLB). The minimization of KL divergence corresponds to the maximization of the VLB, and $q(\mathbf{\Psi})$ equals $P(\mathbf{\Psi}\vert \mathbf{X})$ when KL divergence becomes 0 \parencite{zhang_advances_2019}. To alleviate the complexity in optimizing KL divergence, we assume that the variational distributions can be factorized into $S$ components, which is known as the mean-field assumption on $q(\mathbf{\Psi})$:
\begin{align}
q(\mathbf{\Psi}) = \prod_{s=1}^S q(\psi_s).
\end{align}
Iteratively, we update a set of parameters for each $q(\psi_s)$ in a manner that satisfies $q(\psi_s) \propto \exp(\mathrm{E}_{s'\neq s}[ \log P(\mathbf{X}, \mathbf{\Psi})])$ until the changes of the VLB during iteration achieve the predefined stopping criterion.

In the case of the polytomous-attribute saturated DCM using polytomous-attribute G-matrices, we apply the mean-field assumption to the variational distributions of the model parameters $\mathbf{\Psi}=\{\mathbf{Z}, \mathbf{\Theta}, \bm{\pi}\}$ such that
\begin{align}
q(\mathbf{\Psi}) &= q(\mathbf{Z})q(\mathbf{\Theta}, \bm{\pi}) \nonumber\\
&= \left(\prod_{i=1}^{N}q(\bm{z}_i)\right)\left(\prod_{j=1}^{J}\prod_{l^*=1}^{L_j^*} q(\theta_{jl^*})\right)q(\bm{\pi}).
\end{align}
As mentioned in Section 2.4, the difference between binary- and polytomous-attribute saturated DCMs is in the formulation of a G-matrix and not in the formulation of their joint posteriors. Accordingly, the derivation of the VB algorithm for the binary-attribute saturated DCM elaborated in \textcite{yamaguchi_variational_2021} can be applied directly to the proposed method. \textcolor{black}{The details of the derivation are shown in Supplementary Material A.}

\subsection{Parallelization of the Proposed Algorithm}
VB inference generally consists of two steps: the variational E-step (VE-step) and variational M-step (VM-step). This is because the expectation step (E-step) in an EM algorithm corresponds to updating the variational posteriors of latent variables, and the maximization step (M-step) in an EM algorithm corresponds to updating the variational posteriors of parameters. Hence, the configuration of the parallelized EM algorithm for generalized latent variable models originally developed in \textcite{von_davier_high-performance_2016} is applicable to the proposed method.

After the initialization of the parameters, the parallel-E parallel-M algorithm in von Davier's study (\citeyear{von_davier_high-performance_2016}) starts with parallelizing the E-step using $C$ cores, subdivides examinees into $C$ groups, computes the posteriors of the latent variables given the responses of examinees, and calculates the expected counts of examinees in latent classes in parallel using $C$ worker processes. These computed posteriors are aggregated in a master process. Subsequently, the parallel-M-step proceeds by subdividing items into $C$ groups, computes the gradients of the parameters, and updates the parameters with these computed gradients in parallel using $C$ worker processes. These updated parameters are aggregated in a master process. Finally, a convergence criterion is evaluated using the latent variables and parameters updated in the previous steps. These steps continue until the predefined stopping criterion is achieved.

In the case of the proposed method, we first allocate $C$ cores for parallelization. Then, we subdivide examinees into $C$ groups, update the variational posteriors of the latent variables $q(\bm{z}_i)$ in parallel, and aggregate these results in a master process. Subsequently, the variational posteriors of the mixing proportions $q(\bm{\pi})$ are computed based on the values from the previous step. In the VM-step, we subdivide the items into $C$ groups and update the variational posteriors of the correct-response probability parameters $q(\theta_{jl^*})$ in parallel. These results are aggregated in a master process, and the value of the VLB is computed using the results from the VE- and VM-steps. These steps continue until the predefined stopping criterion is achieved.

\section{Simulation Study}
{\color{black}

\subsection{Simulation Study 1: Performance of the Polytomous-Attribute Saturated DCM with its G-Matrices Using Collapsed Attribute Vectors}
\subsubsection{Simulation Design}
To confirm whether the proposed algorithm can recover the true parameter values, we conducted a simulation study under large-scale conditions. Specifically, we considered a sample size of 10000 or 30000, a number of items of 60 or 120, correlation coefficients among the attributes of .1 or .5, a number of attributes of $K=4$ or $7$, and a number of mastery levels of $M=3$. Additionally, we set the maximum number of attributes measured by one item to $K^*_j=3$ for $K=4$ and to $K^*_j=4$ for $K=7$. With respect to the true Q-matrices, we provide their detailed specifications in Supplementary Material B. The true lowest and highest values of the correct-response probability parameters for each item were randomly generated from the following distributions: $p_j^{\mathrm{lowest}} \sim \mathrm{Uniform}(0.05, 0.25)$ and $p_j^{\mathrm{highest}} \sim \mathrm{Uniform}(0.75, 0.95)$. $p_j^{\mathrm{lowest}}$ corresponds to the guessing probability when an examinee does not master any of the required attributes and $1-p_j^{\mathrm{highest}}$ corresponds to the slip probability when an examinee masters all the required attributes. Given these lowest and highest probability parameters, we monotonically increased the lowest value of the parameters as an examinee acquired the necessary attributes on collapsed attribute vectors to the highest value that corresponds to the correct-response probability when an examinee mastered all the necessary attributes on collapsed attribute vectors for an item. For example, when a \emph{q}-vector for item $j$ is set to $q_j=(1,2,0)$, four unique correct-response probability parameters need to be specified. Accordingly, $p_j^{\mathrm{lowest}}$ is assigned to the correct-response probability given the collapsed attribute vector $\bm{\alpha}_{j1^*}=(0,0)$, meaning that $p(x_{ij}=1|\bm{\alpha}_{j1^*}=(0,0))=p_j^{\mathrm{lowest}}$. Then, we monotonically increase the correct-response probabilities given $\bm{\alpha}_{j2^*}=(1,0)$ and $\bm{\alpha}_{j3^*}=(0,1)$ from $p_j^{\mathrm{lowest}}$ to $p_j^{\mathrm{lowest}}+\frac{1}{2}(p_j^{\mathrm{highest}}-p_j^{\mathrm{lowest}})$. Finally, the correct-response probability given $\bm{\alpha}_{j4^*}=(1,1)$ becomes $p(x_{ij}=1|\bm{\alpha}_{j4^*}=(1,1))=p_j^{\mathrm{highest}}$. Thus, if $p_j^{\mathrm{lowest}}=0.2$ and $p_j^{\mathrm{highest}}=0.8$, we then obtain the following true values of the correct-response probability parameters: $p(x_{ij}=1|\bm{\alpha}_{j1^*}=(0,0))=0.2$; $p(x_{ij}=1|\bm{\alpha}_{j2^*}=(1,0)\text{ or }\bm{\alpha}_{j3^*}=(0,1))=0.5$; and $p(x_{ij}=1|\bm{\alpha}_{j4^*}=(1,1))=0.8$.

Additionally, we conducted a small simulation study for the polytomous-attribute saturated DCM with its G-matrices using reduced attribute vectors. The results are provided in Supplementary Material C.}

To generate the attribute mastery profiles, we applied the following criteria using a multivariate standard normal distribution $\bm{\lambda}_{i} \sim \mathrm{MVM}(\bm{0}, \mathbf{\Sigma})$:
\textcolor{black}{
\begin{align}
\alpha_{ik} = \begin{cases}
M-1 \; \mathrm{if} \;\lambda_{ik} \geq \phi^{-1}(\frac{M-1}{M}) & \\
\vdots &\\
1 \; \mathrm{if} \; \phi^{-1}(\frac{2}{M}) > \lambda_{ik} \geq \phi^{-1}(\frac{1}{M}) & \\
0 \; \mathrm{otherwise}
\end{cases},
\end{align}
where $\bm{\lambda}_{i}$ is a $K$-dimensional vector.} Correlation coefficients among the attributes of \textcolor{black}{$.1$ or $.5$} were assigned in the off-diagonal entries of $\mathbf{\Sigma}$. This generation method was modified from the original criteria used in \textcite{chiu_cluster_2009} and \textcite{liu_data-driven_2012} to produce polytomous attributes. For each artificial dataset, we computed the expected a posteriori (EAP) estimates for the correct-response probability and mixing proportion parameters and the maximum a posteriori (MAP) estimates for the attribute mastery profiles from the corresponding variational posterior distributions. The 100 datasets were simulated for all the conditions.

We assessed the performance of the parameter recovery using the bias and root-mean-square error (RMSE). \textcolor{black}{The bias and RMSE for the correct-response probability parameters $\theta_{jl^*}$ were computed as
\begin{align}
\mathrm{Bias}_{\theta_{jl^*}} = \frac{1}{100}\sum_{t=1}^{T=100} \left(\hat{\theta}_{\mathrm{est}}^{(t)} - \theta_{\mathrm{true}} \right), \\
\mathrm{RMSE}_{\theta_{jl^*}} = \sqrt{ \frac{1}{100}\sum_{t=1}^{T=100} \left(\hat{\theta}_{\mathrm{est}}^{(t)} - \theta_{\mathrm{true}} \right)^2 },
\end{align}
}
where $\theta_{\mathrm{true}}$ denotes the true value of the correct-response probability parameter. We averaged these measures across the related items. \textcolor{black}{In addition, we computed the bias and RMSE for the mixing proportion parameters $\pi_l$ in the same manner as the correct-response probability parameters. Since we could not directly specify the true values of the mixing proportion parameters because of the attribute-profile generation method based on a multivariate standard normal distribution, we first sampled 100 million attribute mastery profiles based on the above multivariate standard normal distribution. Then, the observed proportion of each attribute mastery pattern from this large random sample served as the true value of the corresponding mixing proportion parameter.}

\textcolor{black}{Furthermore, we also calculate the element-wise attribute classification rate (EACR) and pattern-wise attribute classification rate (PACR). These rates are defined as
\begin{align}
EACR_{\alpha_k} &= \frac{1}{100} \sum_{t=1}^{T=100} \frac{1}{N} \sum_{i=1}^N I(\hat{\alpha}_{tik} = \alpha^{true}_{tik}),\\
PACR_{\bm{\alpha}} &= \frac{1}{100} \sum_{t=1}^{T=100} \frac{1}{N} \sum_{i=1}^N I(\hat{\bm{\alpha}}_{ti} = \bm{\alpha}^{true}_{ti}).
\end{align}
}

\subsubsection{Estimation Settings}
We adopted the same settings for the hyperparameters $\bm{\delta}^0$, $\mathbf{A}^0$, and $\mathbf{B}^0$ as in \textcite{yamaguchi_variational_2021}. \textcolor{black}{These hyperparameters control the parametric forms of the prior distributions for the mixing proportion and correct-response probability parameters, and the superscript $0$ indicates that these hyperparameters are relevant to the settings of the prior distributions.} Specifically, a vector of ones was assigned to $\bm{\delta}^0$. \textcolor{black}{This specification yields a non-informative prior, meaning that the prior expectation of the mixing proportion for each attribute mastery pattern is $1/L$.} For $\mathbf{A}^0$ and $\mathbf{B}^0$, we set weakly informative priors such that the prior expectation of the correct-response probability in the pattern with no relevant attributes mastered is below 0.5, and that of the pattern with all relevant attributes mastered is above 0.5. \textcolor{black}{This setting for $\mathbf{A}^0$ and $\mathbf{B}^0$ aims to include the prior information of the monotonicity constraints on the correct-response probabilities into their posterior inference, which helps satisfy the monotonicity constraints in the estimation procedure. We investigated the effects of such a prior in Simulation Study 2.}

\textcolor{black}{Regarding the initial values of the proposed method, the latent indicator variables $z_{il}$} were set to be $1/L$, meaning that the probability of examinee $i$ belonging to pattern $l$ is equivalent across all the patterns. The stopping criteria for the proposed algorithm were specified such that the iteration stopped when the maximum change in the VLB became less than $10^{-4}$ or the number of iterations reached 2000. \textcolor{black}{Finally, the number of cores for parallel computing was set to $C=8$.} 

We wrote all the programs for this study in the Julia programming language \parencite[1.66;][]{bezanson_julia_2017}. You can access the estimation code at the open science framework: \url{https://osf.io/fgn3t/?view_only=7d9df56df89446ef9ed88ed016be7fb4}.

{\color{black}{
\subsubsection{Results}
Table \ref{tab:simulation1_theta_K4} shows the biases and RMSEs for the correct-response probability parameters in the simulation conditions with $K=4$. Owing to a large number of correct-response probability parameters, we summarized their biases and RMSEs in the same manner as \textcite{yamaguchi_variational_2021}, where the biases and RMSEs were averaged according to the number of \textcolor{black}{attributes measured by one item}. We observed that the values of the biases were almost zero in these conditions. Concerning the RMSEs, the increase in sample size lowered their values. Additionally, as the number of \textcolor{black}{attributes measured by one item} increased, the values of RMSEs worsened. These values also deteriorated as the value of the correlation coefficients among the attributes increased under the condition where the number of \textcolor{black}{attributes measured by one item} was two or more. However, the degree of such deterioration was small, and the parameter recovery in the correct-response probability parameters was satisfactory across all the conditions.

\begin{table}[!htbp]
\color{black}
\centering
\caption{Biases and RMSEs for the correct-response probability parameters in the simulation conditions with $K=4$}
\resizebox{\linewidth}{!}{
\begin{tabular}{ccccrccrccr}
\toprule
& & & \multicolumn{8}{c}{Number of Attributes Measured by One Item} \\
\cmidrule{4-11}& & & \multicolumn{2}{c}{1} & & \multicolumn{2}{c}{2} & & \multicolumn{2}{c}{3} \\
\cmidrule{4-5}\cmidrule{7-8}\cmidrule{10-11}Number of Items & Sample Size & Correlation & Bias& RMSE& & Bias& RMSE& & Bias& RMSE \\
\midrule
\multirow{4}[4]{*}{60} & \multirow{2}[1]{*}{10000} & .1& .0000 & .0061 & & -.0002 & .0110 & & .0000 & .0178 \\
& & .5& .0001 & .0060 & & -.0003 & .0134 & & .0000 & .0229 \\
\cmidrule{2-11}& \multirow{2}[1]{*}{30000} & .1& .0000 & .0035 & & -.0001 & .0063 & & .0002 & .0103 \\
& & .5& -.0001 & .0034 & & -.0002 & .0077 & & .0001 & .0132 \\
\midrule
\multirow{4}[4]{*}{120} & \multirow{2}[1]{*}{10000} & .1& -.0001 & .0055 & & -.0001 & .0102 & & -.0002 & .0160 \\
& & .5& -.0001 & .0054 & & -.0001 & .0123 & & -.0001 & .0204 \\
\cmidrule{2-11}& \multirow{2}[1]{*}{30000} & .1& .0001 & .0031 & & .0000 & .0059 & & .0000 & .0092 \\
& & .5& .0000 & .0031 & & .0000 & .0071 & & .0000 & .0118 \\
\bottomrule
\end{tabular}%
}
 \begin{tablenotes}
 \item \footnotesize{\textit{Note.} RMSE = root-mean-square error.}
 \end{tablenotes}
\label{tab:simulation1_theta_K4}%
\end{table}%

\textcolor{black}{
The biases and RMSEs for the mixing proportion parameters are given in Table \ref{tab:simulation1_mixpropK4}. Because of the large number of these parameters, which amounts to $M^K$, we only show the maximum and minimum values of their biases and RMSEs among the $M^K$ parameters. In all the conditions, the absolute values of the largest positive and negative biases were not greater than .0008 and the highest values of the RMSEs were no more than .0033. These results corroborate the sound accuracy of estimating the mixing proportion parameters.}

\begin{table}[!htbp]
\color{black}
\centering
\caption{Maximum and minimum values of the biases and RMSEs for the mixing proportion parameters in the simulation conditions with $K=4$}
\resizebox{\linewidth}{!}{
\begin{tabular}{cccccccc}
\toprule
& & & \multicolumn{5}{c}{Mixing Proportion Parameter} \\
\cmidrule{4-8}& & & \multicolumn{2}{c}{Bias} & & \multicolumn{2}{c}{RMSE} \\
\cmidrule{4-5}\cmidrule{7-8}Number of Items & Sample Size & Correlation & Maximum & Minimum & & Maximum & Minimum \\
\midrule
\multirow{4}[4]{*}{60} & \multirow{2}[1]{*}{10000} & .1& .0003 & -.0005 & & .0020 & .0011 \\
& & .5& .0004 & -.0007 & & .0032 & .0006 \\
\cmidrule{2-8}& \multirow{2}[1]{*}{30000} & .1& .0002 & -.0002 & & .0010 & .0006 \\
& & .5& .0002 & -.0003 & & .0019 & .0003 \\
\midrule
\multirow{4}[4]{*}{120} & \multirow{2}[1]{*}{10000} & .1& .0003 & -.0004 & & .0015 & .0009 \\
& & .5& .0003 & -.0008 & & .0033 & .0005 \\
\cmidrule{2-8}& \multirow{2}[1]{*}{30000} & .1& .0002 & -.0002 & & .0009 & .0004 \\
& & .5& .0002 & -.0003 & & .0017 & .0003 \\
\bottomrule
\end{tabular}%
}
 \begin{tablenotes}
 \item \footnotesize{\textit{Note.} Maximum and minimum values of the biases correspond to the largest positive and negative biases among the $M^K$ mixing proportion parameters, respectively. RMSE = root-mean-square error.}
 \end{tablenotes}
\label{tab:simulation1_mixpropK4}%
\end{table}%

Table \ref{tab:simulation1_attribute_K4} presents the EACRs and PACRs for the attribute profile estimation in the simulation conditions with $K=4$. The EACRs were more than $92\%$ in all the relevant conditions; however, the PACRs in the conditions with $J=60$ were below $80\%$. This relatively unsatisfactory performance in such conditions is attributed to a large number of possible attribute mastery patterns with polytomous attributes and limited number of items. Nonetheless, when the number of items increased to $J=120$, the PACRs climbed above $93\%$.

\begin{table}[!htbp]
\color{black}
\centering
\caption{EACRs and PACRs for the attribute profile estimation in the simulation conditions with $K=4$}
\resizebox{\linewidth}{!}{
\begin{tabular}{cccccccc}
\toprule
& & & \multicolumn{5}{c}{Attribute Classification Rate} \\ 
\cmidrule{4-8} Number of Items & Sample Size & Correlation & $\text{EACR}_{\alpha_1}$ & $\text{EACR}_{\alpha_2}$ & $\text{EACR}_{\alpha_3}$ & $\text{EACR}_{\alpha_4}$ & $\text{PACR}_{\bm{\alpha}}$ \\
\midrule
\multirow{4}[4]{*}{60} & \multirow{2}[1]{*}{10000} & .1& .922& .922& .922& .932& .745 \\
& & .5& .932& .932& .933& .941& .772 \\
\cmidrule{2-8}& \multirow{2}[1]{*}{30000} & .1& .922& .922& .922& .933& .746 \\
& & .5& .933& .932& .933& .942& .773 \\
\midrule
\multirow{4}[4]{*}{120} & \multirow{2}[1]{*}{10000} & .1& .983& .983& .981& .980& .933 \\
& & .5& .985& .985& .983& .983& .940 \\
\cmidrule{2-8}& \multirow{2}[1]{*}{30000} & .1& .983& .983& .981& .980& .933 \\
& & .5& .985& .985& .984& .983& .941 \\
\bottomrule
\end{tabular}%
}
 \begin{tablenotes}
 \item \footnotesize{\textit{Note.} EACR = element-wise attribute classification rate, PACR = pattern-wise attribute classification rate.}
 \end{tablenotes}
\label{tab:simulation1_attribute_K4}%
\end{table}%

\textcolor{black}{
Subsequently, Table \ref{tab:simulation1_theta_K7} shows the biases and RMSEs for the correct-response probability parameters in the simulation conditions with $K=7$. Similar to the conditions with $K=4$, the values of the biases were quite small across all the conditions, where their absolute largest value was only .0022 in the condition with $J=60$, $N=10000$, and the correlation coefficient .5. In contrast to the biases, the relatively large values of the RMSEs were observed when the number of attributes measured by one item was not less than three, where these values ranged from .0103 to .0606. Nonetheless, the values of the RMSEs in other conditions were not more than .0186. The patterns in the effects of the simulation factors were also the same as those with $K=4$, where the increase in the number of attributes measured by one item and value of the correlation coefficients worsened the values of the RMSEs. We also provide the maximum and minimum values of the biases and RMSEs for the mixing proportion parameters in Table \ref{tab:simulation1_mixpropK7}. Similar to those with $K=4$, the small values of the absolute largest positive and negative biases were observed, where these absolute values were less than or equal to .0085. The highest values of RMSEs were also not greater than .0090 in all the conditions, supporting the sound accuracy of parameter estimation in correct-response probability parameters. 
}

\textcolor{black}{
Table \ref{tab:simulation1_attribute_K7} presents the EACRs and PACRs for the attribute profile estimation in the simulation conditions with $K=7$. The EACRs ranged from $79.9\%$ to $88.4\%$ under the conditions with $J=60$, whereas those under the conditions with $J=120$ ranged from $91.6\%$ to $95.4\%$. As the number of items positively contributes to the accuracy of attribute profile estimation, the conditions with a large number of items presented more accurate attribute profile estimation than those with a relatively small number of items. Regarding the PACRs, their values were less than or equal to .372 under the conditions with $J=60$ and .695 under those with $J=120$. Although the increase in the number of items significantly ameliorated the parameter recovery of attributes, the PACRs were significantly worse than those under the conditions with $K=4$ because of the large number of attributes.}

\begin{table}[!htbp]
\color{black}
\centering
\caption{Biases and RMSEs for the correct-response probability parameters in the simulation conditions with $K=7$}
\resizebox{\linewidth}{!}{
\begin{tabular}{cccccccccccccc}
\toprule
& & & \multicolumn{11}{c}{Number of Attributes Measured by One Item} \\
\cmidrule{4-14}& & & \multicolumn{2}{c}{1} & & \multicolumn{2}{c}{2} & & \multicolumn{2}{c}{3} & & \multicolumn{2}{c}{4} \\
\cmidrule{4-5}\cmidrule{7-8}\cmidrule{10-11}\cmidrule{13-14}Number of Items & Sample Size & Correlation & Bias& RMSE& & Bias& RMSE& & Bias& RMSE& & Bias& RMSE \\
\midrule
\multirow{4}[1]{*}{60} & \multirow{2}[1]{*}{10000} & .1& -.0002 & .0104 & & -.0002 & .0141 & & -.0008 & .0234 & & .0003 & .0411 \\
& & .5& -.0012 & .0091 & & -.0022 & .0186 & & -.0006 & .0325 & & -.0014 & .0606 \\
\cmidrule{2-14}& \multirow{2}[1]{*}{30000} & .1& .0001 & .0056 & & .0000 & .0083 & & -.0003 & .0135 & & .0001 & .0238 \\
& & .5& -.0005 & .0051 & & -.0011 & .0113 & & -.0003 & .0187 & & -.0015 & .0375 \\
\midrule
\multirow{4}[1]{*}{120} & \multirow{2}[1]{*}{10000} & .1& -.0002 & .0061 & & -.0002 & .0112 & & -.0005 & .0178 & & -.0001 & .0290 \\
& & .5& -.0004 & .0060 & & -.0007 & .0146 & & -.0004 & .0241 & & -.0004 & .0431 \\
\cmidrule{2-14}& \multirow{2}[1]{*}{30000} & .1& .0001 & .0035 & & .0001 & .0063 & & -.0001 & .0103 & & .0000 & .0167 \\
& & .5& .0000 & .0035 & & -.0001 & .0082 & & .0000 & .0142 & & -.0003 & .0258 \\
\bottomrule
\end{tabular}%
}
 \begin{tablenotes}
 \item \footnotesize{\textit{Note.} RMSE = root-mean-square error.}
 \end{tablenotes}
\label{tab:simulation1_theta_K7}%
\end{table}%

\begin{table}[!htbp]
\color{black}
\centering
\caption{Maximum and minimum values of the biases and RMSEs for the mixing proportion parameters in the simulation conditions with $K=7$}
\resizebox{\linewidth}{!}{
\begin{tabular}{cccccccc}
\toprule
& & & \multicolumn{5}{c}{Mixing Proportion Parameter} \\
\cmidrule{4-8}& & & \multicolumn{2}{c}{Bias} & & \multicolumn{2}{c}{RMSE} \\
\cmidrule{4-5}\cmidrule{7-8}Number of Items & Sample Size & Correlation & Maximum & Minimum & & Maximum & Minimum \\
\midrule
\multirow{4}[1]{*}{60} & \multirow{2}[1]{*}{10000} & .1& .0001 & -.0005 & & .0010 & .0001 \\
& & .5& .0001 & -.0085 & & .0090 & .0001 \\
\cmidrule{2-8}& \multirow{2}[1]{*}{30000} & .1& .0001 & -.0003 & & .0006 & .0001 \\
& & .5& .0001 & -.0033 & & .0036 & .0001 \\
\midrule
\multirow{4}[1]{*}{120} & \multirow{2}[1]{*}{10000} & .1& .0001 & -.0005 & & .0007 & .0001 \\
& & .5& .0001 & -.0083 & & .0085 & .0001 \\
\cmidrule{2-8}& \multirow{2}[1]{*}{30000} & .1& .0000 & -.0001 & & .0004 & .0001 \\
& & .5& .0001 & -.0032 & & .0034 & .0000 \\
\bottomrule
\end{tabular}%
}
 \begin{tablenotes}
 \item \footnotesize{\textit{Note.} Maximum and minimum values of the biases correspond to the largest positive and negative biases among the $M^K$ mixing proportion parameters, respectively. RMSE = root-mean-square error.}
 \end{tablenotes}
\label{tab:simulation1_mixpropK7}%
\end{table}%

\begin{table}[!htbp]
\color{black}
\centering
\caption{EACRs and PACRs for the attribute profile estimation in the simulation conditions with $K=7$}
\resizebox{\linewidth}{!}{
\begin{tabular}{ccccccccccc}
\toprule
& & & \multicolumn{8}{c}{Attribute Classification Rate} \\
\cmidrule{4-11}Number of Items & Sample Size & Correlation & $\text{EACR}_{\alpha_1}$ & $\text{EACR}_{\alpha_2}$ & $\text{EACR}_{\alpha_3}$ & $\text{EACR}_{\alpha_4}$ & $\text{EACR}_{\alpha_5}$ & $\text{EACR}_{\alpha_6}$ & $\text{EACR}_{\alpha_7}$ & $\text{PACR}_{\bm{\alpha}}$ \\
\midrule
\multirow{4}[1]{*}{60} & \multirow{2}[1]{*}{10000} & .1& .799& .823& .862& .858& .821& .830& .813& .287 \\
& & .5& .832& .849& .882& .879& .848& .855& .842& .365 \\
\cmidrule{2-11}& \multirow{2}[1]{*}{30000} & .1& .802& .826& .864& .861& .822& .832& .816& .292 \\
& & .5& .835& .852& .884& .881& .851& .857& .845& .372 \\
\midrule
\multirow{4}[1]{*}{120} & \multirow{2}[1]{*}{10000} & .1& .941& .935& .944& .934& .941& .932& .916& .642 \\
& & .5& .950& .945& .953& .943& .950& .942& .929& .688 \\
\cmidrule{2-11}& \multirow{2}[1]{*}{30000} & .1& .942& .938& .946& .936& .943& .934& .918& .652 \\
& & .5& .951& .946& .954& .945& .951& .944& .931& .695 \\
\bottomrule
\end{tabular}%
}
 \begin{tablenotes}
 \item \footnotesize{\textit{Note.} EACR = element-wise attribute classification rate, PACR = pattern-wise attribute classification rate.}
 \end{tablenotes}
\label{tab:simulation1_attribute_K7}%
\end{table}%

Finally, we present the CPU time and convergence rate for all the conditions in Table \ref{tab:simulation1_cpu}. The proposed method completed the parameter estimation within approximately 50 seconds, and the convergence rates of the estimation were $100\%$ under the conditions with $K=4$. \textcolor{black}{When it comes to $K=7$, the computational speed differs greatly between $J=60$ and $J=120$. Counterintuitively, the CPU time under the conditions with $J=120$ was much shorter than under those with $J=60$. Specifically, the CPU times under the conditions with $J=120$ and $N=30000$ were less than around 1400 to 1500 seconds, whereas these times under those with $J=60$ and $N=30000$ were less than around 3000 to 3100 seconds. The longer CPU time in the conditions with $J=60$ than those with $J=120$ would be attributed to an unsuitable setting for the specified number of cores in the estimation. Lastly, the convergence rates were also $100\%$ under the conditions with $K=7$.} These results generally support the sound stability and computational speed of our method for parameter estimation.

\begin{table}[!htbp]
\color{black}
\centering
\caption{CPU time and convergence rate of the proposed method in Simulation Study 1}
\resizebox{\linewidth}{!}{
\begin{tabular}{ccccc}
\toprule \multicolumn{5}{c}{$K=4$} \\
\toprule
Number of Items & Sample Size & Correlation & CPU Time (seconds) & Convergence Rate \\
\midrule
\multirow{4}[1]{*}{60} & \multirow{2}[1]{*}{10000} & .1& 13.51& $100\%$ \\
& & .5& 12.80& $100\%$ \\
\cmidrule{2-5}& \multirow{2}[1]{*}{30000} & .1& 45.79& $100\%$ \\
& & .5& 41.51& $100\%$ \\
\midrule
\multirow{4}[1]{*}{120} & \multirow{2}[1]{*}{10000} & .1& 13.65& $100\%$ \\
& & .5& 13.92& $100\%$ \\
\cmidrule{2-5}& \multirow{2}[1]{*}{30000} & .1& 43.43& $100\%$ \\
& & .5& 41.48& $100\%$ \\
\bottomrule
& & & &\\
\multicolumn{5}{c}{$K=7$} \\
\toprule
Number of Items & Sample Size & Correlation & CPU Time (seconds) & Convergence Rate \\
\midrule
\multirow{4}[1]{*}{60} & \multirow{2}[1]{*}{10000} & .1& 695.95& $100\%$ \\
& & .5& 682.68& $100\%$ \\
\cmidrule{2-5}& \multirow{2}[1]{*}{30000} & .1& 3058.70& $100\%$ \\
& & .5& 3011.15& $100\%$ \\
\midrule
\multirow{4}[1]{*}{120} & \multirow{2}[1]{*}{10000} & .1& 591.55& $100\%$ \\
& & .5& 559.19& $100\%$ \\
\cmidrule{2-5}& \multirow{2}[1]{*}{30000} & .1& 1506.23& $100\%$ \\
& & .5& 1396.65& $100\%$ \\
\bottomrule
\end{tabular}%
}
\label{tab:simulation1_cpu}%
\end{table}%

\clearpage
\subsection{Simulation Study 2: Effects of Prior Information on Monotonicity Constraints}
One concern about the proposed method is that the monotonicity constraints on correct-response probabilities cannot be directly incorporated into the estimation procedure. Since these constraints are essential for the interpretable clustering of respondents based on attribute mastery profiles and meaningful item parameter estimates \parencite{yamaguchi_gibbs_2021}, the issue of the constraints must be addressed in the proposed method. To cope with this issue, we employed the weakly informative priors of the monotonicity constraints in Simulation Study 1. The same approach was also adopted in \textcite{yamaguchi_variational_2020}'s study of the VB estimation for the DINA model, where they specified the priors for the slip and guessing parameters in such a manner that the prior expectations of both parameters were less than .50. In this simulation study, we assessed the effects of such prior information on monotonicity in the correct-response probability parameters. Specifically, we compared the estimation performance between the proposed methods equipped with and without these prior specifications to investigate how such priors can stabilize estimating the correct-response probability parameters with monotonicity.

Regarding the simulation condition, we used the same datasets under the setting $K=4$, $J=60$, and $N=10000\text{ or } 30000$; the correlation coefficient among the attributes is .1 as in Simulation Study 1. Additionally, for the proposed method with non-informative priors, we assigned $a^0_{jl^*}=1$ and $b^0_{jl^*}=1$ to the priors on all the correct-response probability parameters, which gives no prior information to posterior inference. All the other specifications were set to be the same as those in Simulation Study 1. 

\subsubsection{Results}
The results of Simulation Study 2 are provided in Table \ref{tab:simulation3}. As shown in the table, the proposed method with non-informative priors performed the parameter estimation as accurately as that with the weakly informative priors of the monotonicity constraints. \textcolor{black}{Additionally, we found that the methods with both types of priors did not violate monotonicity in the correct-response probability parameters under any of the relevant conditions.}

\begin{table}[!htbp]
\color{black}
\centering
\caption{The results in Simulation Study 2 ($K=4$, correlation coefficient .1)}
\resizebox{\linewidth}{!}{
\begin{tabular}{ccccccccccc}
\toprule
\multicolumn{11}{c}{Biases and RMSEs for the correct-response probability parameters}\\
\toprule
& & & \multicolumn{8}{c}{Number of Attributes Measured by One Item} \\
\cmidrule{4-11}& & & \multicolumn{2}{c}{1} & & \multicolumn{2}{c}{2} & & \multicolumn{2}{c}{3} \\
\cmidrule{4-5}\cmidrule{7-8}\cmidrule{10-11}Prior Settings & Number of Items & Sample Size & Bias& RMSE& & Bias& RMSE& & Bias& RMSE \\
\midrule
\multirow{2}[1]{*}{Weakly Informative Prior} & \multirow{2}[1]{*}{60} & 10000 & .0000 & .0061 & & -.0002 & .0110 & & .0000 & .0178 \\
& & 30000 & .0000 & .0035 & & -.0001 & .0063 & & .0002 & .0103 \\
\midrule
\multirow{2}[1]{*}{Non-Informative Prior} & \multirow{2}[1]{*}{60} & 10000 & .0002 & .0061 & & .0000 & .0110 & & .0000 & .0179 \\
& & 30000 & .0001 & .0035 & & .0000 & .0063 & & .0002 & .0103 \\
\bottomrule \\
\\
\end{tabular}%
}

\resizebox{\linewidth}{!}{
\begin{tabular}{cccccccc}
\multicolumn{8}{c}{Maximum and minimum values of the biases and RMSEs for the mixing proportion parameters}\\
\toprule
& & & \multicolumn{5}{c}{Mixing Proportion Parameter} \\
\cmidrule{4-8}& & & \multicolumn{2}{c}{Bias} & & \multicolumn{2}{c}{RMSE} \\
\cmidrule{4-5}\cmidrule{7-8}Prior Settings & Number of Items & Sample Size & Maximum & Minimum & & Maximum & Minimum \\
\midrule
\multirow{2}[1]{*}{Weakly Informative Prior} & \multirow{2}[1]{*}{60} & 10000 & .0003 & -.0005 & & .0020 & .0011 \\
& & 30000 & .0002 & -.0002 & & .0010 & .0006 \\
\midrule
\multirow{2}[1]{*}{Non-Informative Prior} & \multirow{2}[1]{*}{60} & 10000 & .0003 & -.0005 & & .0020 & .0011 \\
& & 30000 & .0002 & -.0002 & & .0010 & .0006 \\
\bottomrule \\
\\
\end{tabular}%
}

\resizebox{\linewidth}{!}{
\begin{tabular}{cccccccc}
\multicolumn{8}{c}{EACRs and PACRs for the attribute profile estimation}\\
\toprule
& & & \multicolumn{5}{c}{Attribute Classification Rate} \\
\cmidrule{4-8}Prior Settings & Number of Items & Sample Size & $\text{EACR}_{\alpha_1}$ & $\text{EACR}_{\alpha_2}$ & $\text{EACR}_{\alpha_3}$ & $\text{EACR}_{\alpha_4}$ & $\text{PACR}_{\bm{\alpha}}$ \\
\midrule \multirow{2}[1]{*}{Weakly Informative Prior} & \multirow{2}[1]{*}{60} & 10000 & .922& .922& .922& .932& .745 \\
& & 30000 & .922& .922& .922& .933& .746 \\
\midrule
\multirow{2}[1]{*}{Non-Informative Prior} & \multirow{2}[1]{*}{60} & 10000 & .922& .922& .922& .932& .745 \\
& & 30000 & .922& .922& .922& .933& .746 \\
\bottomrule
\end{tabular}%
}
 \begin{tablenotes}
 \item \footnotesize{\textit{Note.} Maximum and minimum values of the biases correspond to the largest positive and negative biases among the $M^K$ mixing proportion parameters, respectively. RMSE = root-mean-square error, EACR = element-wise attribute classification rate, PACR = pattern-wise attribute classification rate.}
 \end{tablenotes}
\label{tab:simulation3}%
\end{table}%

}}

\clearpage
\section{Empirical Study}
We conducted an empirical study to show the utility of the proposed method by comparing it with the performance of an MCMC estimation. The dataset from the standardized achievement test for Grade 9 Mathematics garnered in 2019 by TOKYO SHOSEKI CO., LTD. was used for the empirical evaluation. The contents of this examination target four aspects of mathematics: numbers and algebraic expressions, geometrical figures, functions, and making use of data. Examples of items modified from the original ones include ``simplify: $(-12)xy^2 \div 6xy \times (-4xy)$'' and ``solve for $a$ in the following equation: $4a -7 = 8a -9$.'' The sample size is 21,888, and the number of items is 34.

Regarding the attribute specification, the second author identified three binary and polytomous attributes for this test based on perspectives from cognitive and educational psychology. This Q-matrix was validated by two educational psychologists, and its inconsistencies in the validation process were discussed between them. The first attribute is binary and this represents a computational skill. The second attribute is polytomous with the levels of $M=3$ and this represents the (1) comprehension of the procedure and (2) conceptual understanding of the procedure. The third attribute is binary and this represents the comprehension of the terminology. The detailed specification of the Q-matrix for this examination is shown in Table \ref{tab:realdataQ}.
\begin{table}[!htbp]
\centering
\caption{The specification of the Q-matrix for the empirical study}
\resizebox{\linewidth}{!}{
\begin{tabular}{cccc}
\toprule
& \multicolumn{3}{c}{Q-matrix} \\
\cmidrule{2-4}& \multicolumn{1}{c}{Attribute 1 (binary)} & \multicolumn{1}{c}{Attribute 2 (polytomous)} & \multicolumn{1}{c}{Attribute 3 (binary)} \\
Item ID & \multicolumn{1}{c}{\begin{tabular}{c}Computational \\Skill (1)\end{tabular}} & \multicolumn{1}{c}{\begin{tabular}{c}Comprehension of Procedure (1) and \\ Conceptual Understanding of Procedure (2) \end{tabular} } & \multicolumn{1}{c}{\begin{tabular}{c}Comprehension of\\ Terminology (1)\end{tabular}} \\
\midrule
1 & 1 & 0 & 0 \\
2 & 1 & 0 & 0 \\
3 & 1 & 0 & 0 \\
4 & 1 & 0 & 0 \\
5 & 1 & 0 & 0 \\
6 & 1 & 0 & 0 \\
7 & 1 & 0 & 0 \\
8 & 0 & 1 & 1 \\
9 & 0 & 1 & 0 \\
10& 0 & 1 & 1 \\
11& 1 & 1 & 0 \\
12& 0 & 2 & 0 \\
13& 0 & 1 & 0 \\
14& 1 & 1 & 1 \\
15& 1 & 1 & 1 \\
16& 1 & 1 & 1 \\
17& 1 & 1 & 0 \\
18& 1 & 1 & 0 \\
19& 1 & 1 & 0 \\
20& 1 & 1 & 0 \\
21& 1 & 1 & 0 \\
22& 1 & 1 & 0 \\
23& 1 & 1 & 0 \\
24& 0 & 0 & 1 \\
25& 0 & 2 & 0 \\
26& 0 & 0 & 1 \\
27& 1 & 1 & 0 \\
28& 1 & 1 & 1 \\
29& 1 & 1 & 1 \\
30& 1 & 1 & 1 \\
31& 0 & 1 & 1 \\
32& 0 & 1 & 0 \\
33& 0 & 2 & 0 \\
34& 0 & 2 & 0 \\
\bottomrule
\end{tabular}%
}
\label{tab:realdataQ}%
\end{table}%

Concerning the MCMC estimation, we developed the Gibbs sampler for the polytomous-attribute saturated DCM using its G-matrices based on collapsed attribute vectors. The specifications of the hyperparameters were set to be the same as those in Simulation Study 1. When implementing this Gibbs sampler, we run three chains of 5000 iterations with 2000 as the burn-in. We assessed the convergence of MCMC chains via the convergence diagnostics called the rank-normalized split-$\hat{R}$ proposed in \textcite{vehtari_rank-normalization_2021} and confirmed that all the parameters satisfied $\hat{R}<1.05$. Additionally, we parallelized generating MCMC chains using three cores. For the proposed method, we assigned the same specifications as in Simulation Study 1 and \textcolor{black}{set the same number of cores as the Gibbs sampler, which corresponds to $C=3$.}

\textcolor{black}{First, the computational times of the Gibbs sampler and the proposed method with parallel computing were 777.21 and 12.45 seconds, respectively. As expected, the parallelized VB algorithm presented a superior computational speed over} the parallelized Gibbs sampler, and the computational time of our method was about \textcolor{black}{60} times faster than that of the Gibbs sampler. \textcolor{black}{With respect to the effect of parallel computing, the computational time of the proposed method without parallel computing was 30.30 seconds. As the number of attributes is three, the parallelized proposed method reduces the computational time without parallel computing by only about $60\%$. Nonetheless, the parallelization would yield much better computational efficiency as the attribute dimensionality rises.}

\begin{figure}[!htbp]
 \begin{center}
 \includegraphics[width=\linewidth, keepaspectratio]{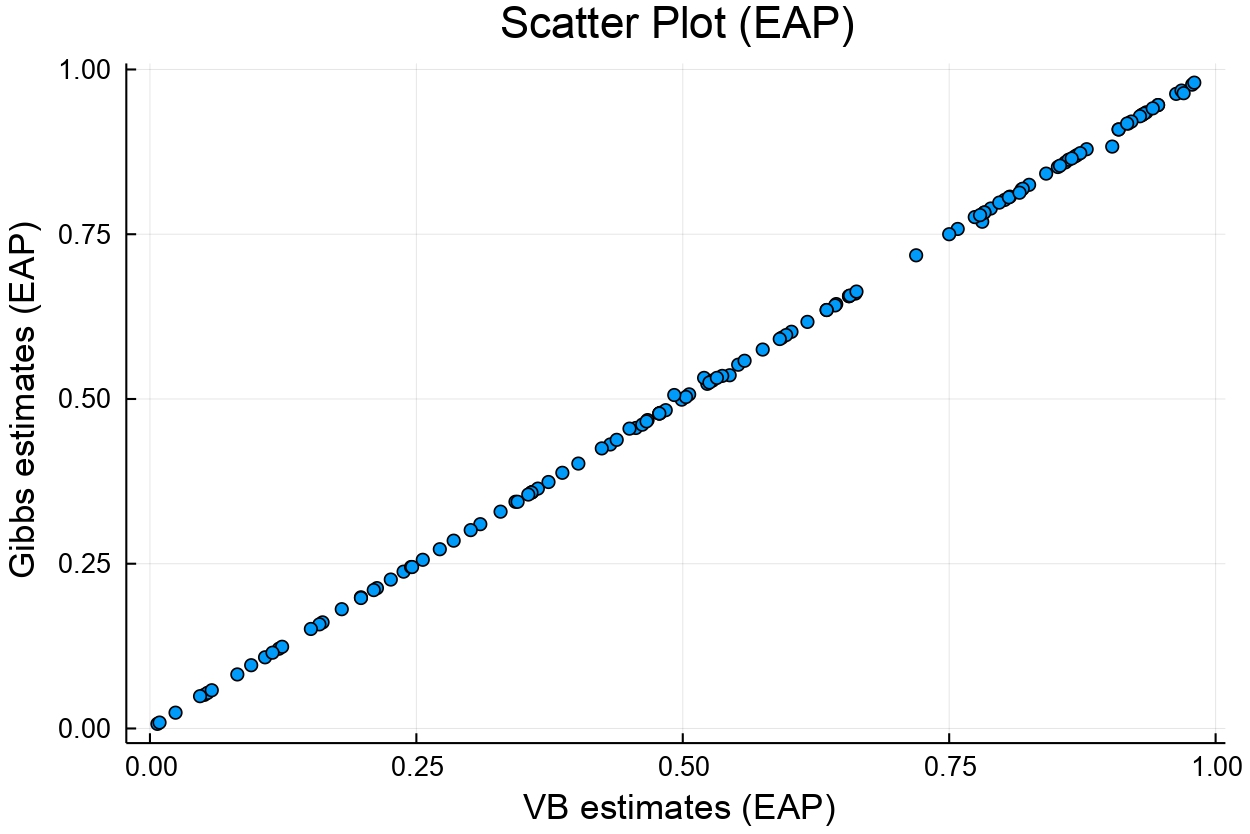}
 \end{center}
 \caption{The scatter plot of the EAP estimates for the correct-response probability parameters from the proposed method and Gibbs sampler.}
 \label{fig:scatter_theta_eap}
\end{figure}

\begin{figure}[!htbp]
 \begin{center}
 \includegraphics[width=\linewidth, keepaspectratio]{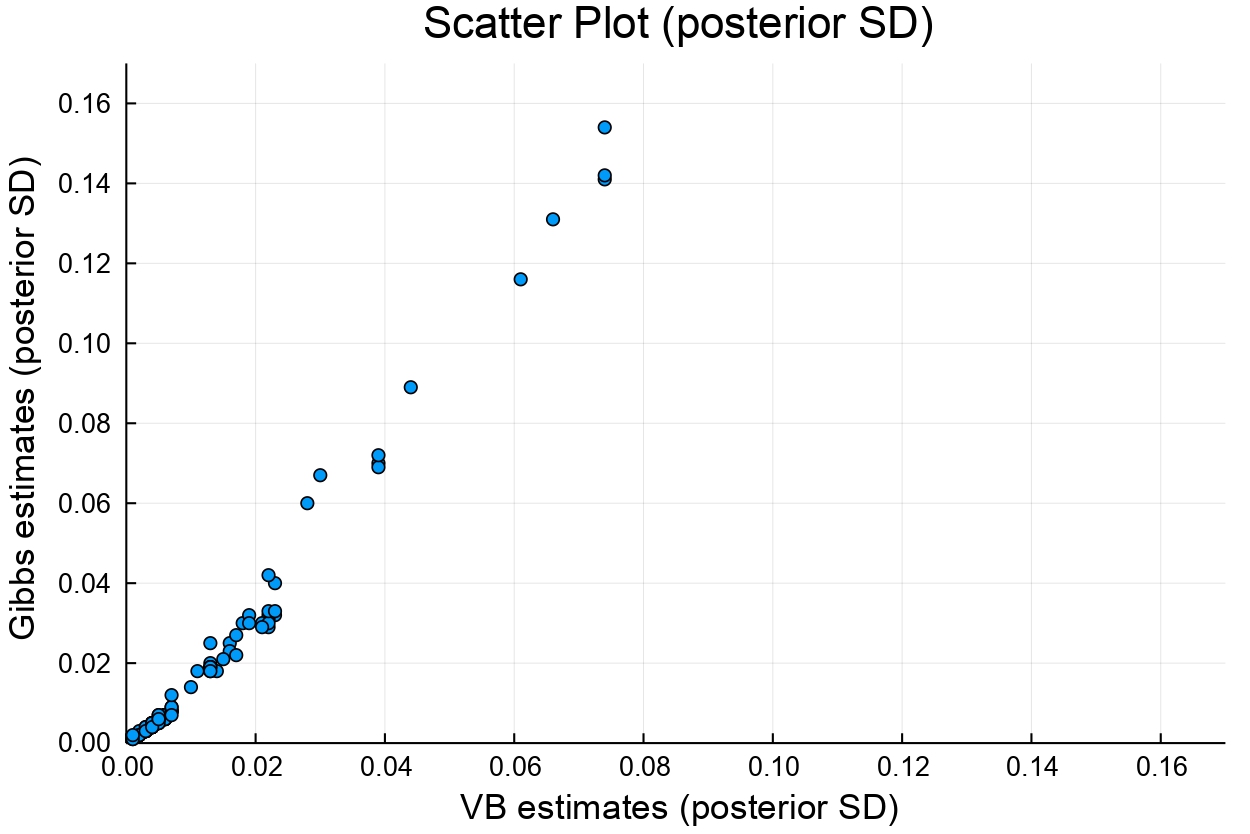}
 \end{center}
 \caption{The scatter plot of the posterior standard deviation for the correct-response probability parameters from the proposed method and Gibbs sampler.}
\label{fig:scatter_theta_psd}
\end{figure}

\textcolor{black}{Second, we show the results of the parameter estimation for the correct-response probability parameters. Owing to space limitations, we provide their estimates of EAP and posterior standard deviation (SD) for each item in Supplementary Material D. Here, we present the scatter plot of the EAP and posterior SD for the parameter estimates. Figure \ref{fig:scatter_theta_eap} shows that the EAP estimates of the correct-response probability parameters from our method are closely aligned with those from the Gibbs sampler. Subsequently, Figure \ref{fig:scatter_theta_psd} shows their posterior SD estimates. In general, the estimates from our method were close to those from the Gibbs sampler. However, our method moderately underestimated the posterior SDs compared with the Gibbs sampler for the items for which the corresponding MCMC-estimated posterior SDs were relatively large. This underestimation of posterior variance is a well-known tendency of VB inference \parencite{bishop_pattern_2006}. Nonetheless, the largest absolute difference in the EAP estimates for the correct-response probability parameters between these two estimation methods was .0180, which coincides with $\theta_{jl^*}$ with $j=30$ and $l^*=4$, while that for the posterior SD estimates was .0730, which corresponds to $\theta_{jl^*}$ with $j=28$ and $l^*=4$. These results indicate} that the estimated values of the parameters from the proposed method were highly consistent with those from the Gibbs sampler.

\textcolor{black}{Third}, Table \ref{tab:mixingprop} presents the EAP and posterior SD estimates for the mixing proportion parameters. Similar to the correct-response probability parameters, the EAP and posterior SD estimates from our method were closely aligned with those from the Gibbs sampler. The largest absolute difference in the EAP estimates \textcolor{black}{for the mixing proportion parameters was .00024 and that in the posterior SD estimate was .00216. Both these values correspond to the mixing proportion parameter of the attribute mastery class with $\bm{\alpha}=(1,1,1)$.} These results indicate that our method can estimate model parameters as accurately as the Gibbs sampler.

\textcolor{black}{Lastly, we computed the percentage of the matched entries in the estimated attribute mastery patterns between the proposed method and Gibbs sampler. The element-wise matched ratios for the attributes were .9993 (A1), .9989 (A2), and .9990 (A3) and the pattern-wise matched ratio was .9981. This result also corroborates the high similarity of the parameter estimates between these two methods.}

\begin{table}[!htbp]
\centering
\caption{The estimates of EAP and posterior standard deviation for the mixing proportion parameters}
\resizebox{\linewidth}{!}{
\begin{tabular}{rlrrrrrrrrrrrr}
\toprule
& & \multicolumn{12}{c}{Attribute Mastery Patterns} \\
\cmidrule{3-14}& & \multicolumn{1}{c}{000} & \multicolumn{1}{c}{001} & \multicolumn{1}{c}{010} & \multicolumn{1}{c}{011} & \multicolumn{1}{c}{020} & \multicolumn{1}{c}{021} & \multicolumn{1}{c}{100} & \multicolumn{1}{c}{101} & \multicolumn{1}{c}{110} & \multicolumn{1}{c}{111} & \multicolumn{1}{c}{120} & \multicolumn{1}{c}{121} \\
\midrule
 \multicolumn{1}{l}{Gibbs} & Mean& .259& .007& .018& .001& .004& .001& .064& .232& .020& .022& .002& .368 \\
 & SD & .003& .001& .002& .000& .001& .001& .003& .004& .002& .003& .001& .005 \\
 \multicolumn{1}{l}{VB} & Mean& .259& .007& .018& .001& .004& .001& .064& .232& .020& .022& .002& .368 \\
 & SD & .003& .001& .001& .000& .000& .000& .002& .003& .001& .001& .000& .003 \\
\bottomrule
\end{tabular}%
}
\label{tab:mixingprop}
\end{table}%

\section{Conclusion and Discussion}

To evaluate the performance of the proposed method, we conducted a Monte Carlo simulation under the simulation design with \textcolor{black}{large-scale} conditions. Three main findings were observed in \textcolor{black}{Simulation Study 1}. First, the biases of the correct-response probability parameters were small across all the conditions. Second, the following general tendencies were found: (1) sample size reduced the values of the RMSEs, (2) as \textcolor{black}{the number of attributes measured by one item} increased, the values of the RMSEs worsened, and (3) as the value of the correlation coefficients among the attributes increased, the RMSEs deteriorated when the number of \textcolor{black}{attributes measured by one item} was greater than or equal to two. Finally, the relatively high values of the RMSEs were observed when the number of \textcolor{black}{attributes measured by one item increased. However, the effect of the factors deteriorating the parameter estimation was small, and} estimation accuracy was generally satisfactory under a wide range of conditions. \textcolor{black}{In addition, we conducted a simulation to investigate the effect of the weakly informative priors of monotonicity constraints in Simulation Study 2. The findings from this investigation show that the estimation methods with the weakly informative and non-informative priors produced identical parameter estimates. They also did not yield estimates of the correct-response probability parameters without monotonicity. These results suggest that, regardless of the prior specifications, the proposed method can stably estimate monotonic correct-response probability parameters when the true values of these probability parameters satisfy monotonicity.}

Through the empirical study, we observed that our method presented computational superiority over the Gibbs sampler; in fact, the computational time of our method was about \textcolor{black}{60} times faster than that of the Gibbs sampler. Moreover, our method estimated the model parameters using only a small amount of memory in the computer during the estimation, whereas the Gibbs sampler presented substantial memory usage for its estimation. Accordingly, our method was significantly more computationally efficient than the Gibbs sampler in terms of both computational time and memory usage. \textcolor{black}{Additionally, parallelization moderately reduced the computational time of our method compared with that without parallel computing. Although the effect of parallelization was relatively modest, its degree would increase when the numbers of samples and attributes rise.} Regarding the estimates of EAP and posterior SD for the correct-response probability and mixing proportion parameters, both estimation methods produced highly similar estimates. \textcolor{black}{The high similarity of estimates was also found in the attribute profile estimation.} Since the observed convergence diagnostics corroborated that the MCMC samples from the Gibbs sampler accurately constructed targeted posteriors, this result provides strong evidence of the sound accuracy of our method in variational approximation to targeted posteriors. \textcolor{black}{However, the attribute dimensionality of the Q-matrix in the empirical study is three, and only one attribute is polytomous. Accordingly, this simple specification of the Q-matrix may contribute to the high similarity of the parameter estimates between the VB algorithm and the Gibbs sampler. Thus, the proposed method may provide parameter estimates differing from those of a Gibbs sampler when the dimensionality and polytomy of attributes become larger.}

\textcolor{black}{The limitations of the proposed method are twofold. First, although our method incorporates the weakly informative priors of the monotonicity constraints into the posterior inference for the correct-response probability parameters, it does not ensure that these probability parameter estimates always hold monotonicity. Rather, monotonicity can still be violated even when we use such informative priors. Hence, a VB algorithm that can strictly enforce the monotonicity constraints in its estimation procedure should be developed in future research.}

Second, the susceptibility of the parameter estimates to the initial values can be observed in the estimation. \textcite{yamaguchi_variational_2021} reported in their simulation study that the choices of initial values hardly influenced the biases of the parameter estimates but did influence their RMSEs. This varying effect of the initial values on the parameter estimation can also be expected in the proposed method because the estimation procedure of our method is comparable with the binary-attribute saturated DCM except for the implementation of parallel computing. This issue can be easily addressed by introducing simulated annealing in the optimization process of the VLB, such as deterministic annealing variational Bayes \parencite{katahira_deterministic_2008}. However, in practice, we would recommend implementing the proposed method with different initial values multiple times and choosing the result with the maximum value of the VLB.

Concerning future directions of this study, it is practically important to develop a scalable Q-matrix estimation method for polytomous-attribute DCMs. \textcolor{black}{Although a Q-matrix is assumed to be known in the proposed method, it is commonly provisional or unknown in real-world settings. To leverage the utility of polytomous-attribute DCMs in such settings}, estimation methods to learn a polytomous-attribute Q-matrix from data are required to conduct accurate diagnosis because misspecifications of a Q-matrix are known to be detrimental to parameter estimation \parencite{rupp_effects_2008,kunina-habenicht_impact_2012}. Although accurate Q-matrix estimation is computationally demanding or difficult even in binary-attribute DCMs \parencite[e.g.,][]{xu_identifying_2018,chung_gibbs_2019, liu_constrained_2020}, it is a promising direction to extend existing Q-matrix estimation methods in binary-attribute DCMs to polytomous-attribute DCMs or to develop a VB-based Q-matrix estimation method for polytomous-attribute DCMs.

In conclusion, the present study developed a parallelized VB algorithm for a polytomous-attribute saturated DCM and demonstrated its utility through simulation and empirical studies. The proposed method serves as a scalable psychometric method that can be applicable to \textcolor{black}{big educational data}.

\section*{Statements and Declarations}
\subsection*{Acknowledgements}
The authors would like to express their sincere gratitude to TOKYO SHOSEKI CO., LTD. for generously sharing the data used in our empirical study, to Dr. Haruhiko Mitsunaga for 
his support under JSPS KAKENHI (B) Grant Number JP20H01720, and to Dr. Kazuhiro Yamaguchi for helpful discussions and comments on the proposed method. A preliminary report of this study was presented at the World Meeting of the International Society for Bayesian Analysis 2021.

\subsection*{Funding}
This work was supported by JSPS KAKENHI (Grant Numbers 19H00616, 21H00936, and 20H01720).

\subsection*{Conflict of Interest}
The authors declare that they have no conflict of interest.

\section*{References}
\printbibliography[heading=none]

\clearpage
\section*{Supplementary Materials}

\subsection*{Supplementary Material A\footnote{The derivation in Supplementary Material A is primarily an adaptation of that of the binary-attribute DCMs shown in Yamaguchi and Okada (2021).}}

In Supplementary Material A, we provide the derivation of the conditional variational posterior for each of the parameters and the lower bound of the log marginal likelihood (variational lower bound; VLB) in the polytomous-attribute DCM.

First, the log joint posterior of the polytomous-attribute DCM is defined as
\begin{align}
&\log{P(\mathbf{Z}, \bm{\Theta}, \bm{\pi} \vert \mathbf{X}, \mathbf{G}^{\mathrm{poly}}, \mathbf{Q}, \bm{\delta}^0, \mathbf{A}^{0}, \mathbf{B}^{0})} \nonumber\\
&\quad \propto \log{P(\mathbf{X} \vert \mathbf{Z},\bm{\Theta},\mathbf{G}^{\mathrm{poly}}, \mathbf{Q})P(\mathbf{Z} \vert \bm{\pi})P(\bm{\pi} \vert \bm{\delta}^0)P(\bm{\Theta} \vert \mathbf{A}^{0}, \mathbf{B}^{0})} \nonumber\\
&\quad =\sum_{i=1}^{N}\sum_{j=1}^{J}\sum_{l^*=1}^{L^*_j} z_{ijl^*}\Biggl\{x_{ij}\log\theta_{jl^*} + (1-x_{ij})\log(1-\theta_{jl^*}) \Biggl\} \nonumber\\
&\qquad + \sum_{i=1}^{N}\sum_{l=1}^{L}z_{il}\log\pi_l + \sum_{l=1}^{L}(\delta_l^0-1)\log\pi_l\nonumber\\
&\qquad + \sum_{j=1}^{J}\sum_{l^*=1}^{L_j^*}\Biggl\{(a_{jl^*}^0-1)\log\theta_{jl^*} + (b_{jl^*}^0-1)\log(1-\theta_{jl^*})\Biggl\}\nonumber\\
&\quad = \sum_{i=1}^{N}\sum_{j=1}^{J}\sum_{l^*=1}^{L^*_j} \Biggl(\sum_{l=1}^{L}g_{jl^*l}z_{il}\Biggl)\Biggl\{x_{ij}\log\theta_{jl^*} + (1-x_{ij})\log(1-\theta_{jl^*}) \Biggl\} \nonumber\\
&\qquad \Bigl(\because z_{ijl^*}=\sum_{l=1}^{L}g_{jl^*l}z_{il}\Bigl)\nonumber\\
&\qquad + \sum_{i=1}^{N}\sum_{l=1}^{L}z_{il}\log\pi_l + \sum_{l=1}^{L}(\delta_l^0-1)\log\pi_l\nonumber\\
&\qquad + \sum_{j=1}^{J}\sum_{l^*=1}^{L_j^*}\Biggl\{(a_{jl^*}^0-1)\log\theta_{jl^*} + (b_{jl^*}^0-1)\log(1-\theta_{jl^*})\Biggl\}.
\end{align}
Next, under the mean-field assumption of the variational distribution such that $q(\bm{\Psi})=q(\mathbf{Z})q(\bm{\Theta},\bm{\pi})=\prod_{i=1}^{N}q(\bm{z}_i)\prod_{j=1}^J\prod_{l^*=1}^{L_j^*}q(\theta_{jl^*})q(\bm{\pi})$, where $\bm{\Psi}=\{\mathbf{Z}, \bm{\Theta},\bm{\pi}\}$, we derive the conditional variational posterior for each of the parameters in the polytomous-attribute saturated DCM.

\newpage
\subsubsection*{A1. Inference of $\bm{z}_i$}
By applying the formula for updating the parameters of the variational posteriors in Section 2.5, the log variational posterior $\log q(\bm{z}_i)$ is obtained as
\begin{align}
& \log q(\bm{z}_i) \nonumber\\
& \quad = \mathrm{E}_{q(\bm{\Theta}, \bm{\pi})}\Biggl[ \sum_{j=1}^{J}\sum_{l^*=1}^{L^*_j} (\sum_{l=1}^{L}g_{jl^*l}z_{il})\biggl\{x_{ij}\log\theta_{jl^*} + (1-x_{ij})\log(1-\theta_{jl^*}) \biggl\} + \sum_{l=1}^{L}z_{il}\log\pi_l \Biggl] \nonumber\\
& \qquad + \text{ const} \nonumber\\
& \quad = \mathrm{E}_{q(\bm{\Theta}, \bm{\pi})}\Biggl[
\sum_{l=1}^{L}z_{il}\Biggl(\sum_{j=1}^{J}\sum_{l^*=1}^{L^*_j}g_{jl^*l}\biggl\{x_{ij}\log\theta_{jl^*} + (1-x_{ij})\log(1-\theta_{jl^*}) \biggl\} + \log\pi_l \Biggl) \Biggl] \nonumber\\
& \qquad + \text{ const} \nonumber\\
& \quad = 
\sum_{l=1}^{L}z_{il}\Biggl(\sum_{j=1}^{J}\sum_{l^*=1}^{L^*_j}g_{jl^*l}\biggl\{x_{ij}\mathrm{E}_{q(\theta_{il^*})}[\log\theta_{jl^*}] + (1-x_{ij})\mathrm{E}_{q(\theta_{il^*})}[\log(1-\theta_{jl^*})] \biggl\} \nonumber\\
& \qquad + \mathrm{E}_{q(\bm{\pi})}[\log\pi_l] \Biggl) + \text{ const}\nonumber\\
& \quad = \sum_{l=1}^{L}z_{il}\log\rho_{il} + \text{ const},
\end{align}
where 
\begin{align}
\log\rho_{il}&=\sum_{j=1}^{J}\sum_{l^*=1}^{L^*_j}g_{jl^*l}\biggl\{x_{ij}\mathrm{E}_{q(\theta_{jl^*})}[\log\theta_{jl^*}] + (1-x_{ij})\mathrm{E}_{q(\theta_{jl^*})}[\log(1-\theta_{jl^*})] \biggl\} \nonumber\\ 
& \quad + \mathrm{E}_{q(\bm{\pi})}[\log\pi_l].
\end{align}
Thus, the conditional variational posterior of $\bm{z}_i$ is expressed as the following categorical distribution:
\begin{align}
q(\bm{z}_i) \propto \prod_{l=1}^{L}r_{il}^{z_{il}},
\end{align}
where $r_{il}=\frac{\rho_{il}}{\sum_{l=1}^{L}\rho_{il}}$.

\subsubsection*{A2. Inference of $\bm{\pi}$}
Similarly, the log variational posterior $\log q(\bm{\pi})$ is obtained as
\begin{align}
\log q(\bm{\pi}) & = \mathrm{E}_{q(\mathbf{Z})q( \bm{\Theta})}\Biggl[ \sum_{i=1}^{N}\sum_{l=1}^{L}z_{il}\log\pi_l + \sum_{l=1}^{L}(\delta_l^0-1)\log\pi_l \Biggl] + \text{ const}\nonumber\\
& = \sum_{l=1}^{L}\log\pi_l \Biggl(\Bigl\{\sum_{i=1}^{N}\mathrm{E}_{q(\bm{z}_i)}[z_{il}]\Bigl\} + \delta_l^0-1\Biggl) + \text{ const}.
\end{align}
This is equivalent to the following Dirichlet distribution:
\begin{align}
q(\bm{\pi}) \propto \prod_{l=1}^{L}\pi_l^{\delta^*_l - 1},
\end{align}
where $\delta^*_l = \Bigl\{\sum_{i=1}^{N}\mathrm{E}_{q(\bm{z}_i)}z_{il}\Bigl\} + \delta_l^0$.

\subsubsection*{A3. Inference of $\theta_{jl^*}$}
Finally, the log variational posterior $\log q(\theta_{jl})$ is obtained as 
\begin{align}
\log q(\theta_{jl}) & = \mathrm{E}_{q(\mathbf{Z})q(\bm{\pi})}\Biggl[
\sum_{i=1}^{N} \Biggl(\sum_{l=1}^{L}g_{jl^*l}z_{il}\Biggl)\Biggl\{x_{ij}\log\theta_{jl^*} + (1-x_{ij})\log(1-\theta_{jl^*}) \Biggl\} \nonumber\\
& \quad + (a_{jl^*}^0-1)\log\theta_{jl^*} + (b_{jl^*}^0-1)\log(1-\theta_{jl^*}) \Biggl] + \text{ const} \nonumber\\
& = \Biggl(\sum_{i=1}^{N}\sum_{l=1}^{L} g_{jl^*l}\mathrm{E}_{q(\bm{z}_i)}[z_{il}]x_{ij} + a_{jl^*}^0-1 \Biggl)\log\theta_{jl^*} \nonumber\\
& \quad +\Biggl(\sum_{i=1}^{N}\sum_{l=1}^{L} g_{jl^*l}\mathrm{E}_{q(\bm{z}_i)}[z_{il}](1-x_{ij}) + b_{jl^*}^0-1 \Biggl)\log(1-\theta_{jl^*}) + \text{ const}.
\end{align}
The right-hand side of the above equation is equivalent to the following Beta distribution:
\begin{align}
q(\theta_{jl^*}) \propto \theta_{jl^*}^{a^*_{jl^*}-1}(1-\theta_{jl^*})^{b^*_{jl^*}-1},
\end{align}
where
\begin{align}
\begin{cases}
a^*_{jl^*} = \sum_{i=1}^{N}\sum_{l=1}^{L} g_{jl^*l}\mathrm{E}_{q(\bm{z}_i)}[z_{il}]x_{ij} + a_{jl^*}^0 & \\
b^*_{jl^*} = \sum_{i=1}^{N}\sum_{l=1}^{L} g_{jl^*l}\mathrm{E}_{q(\bm{z}_i)}[z_{il}](1-x_{ij}) + b_{jl^*}^0
\end{cases}.
\end{align}

\newpage
\subsubsection*{A4. Deriving the Variational Lower Bound}
We provide the derivation of the VLB of the log marginal likelihood for the polytomous-attribute saturated DCM with G-matrices using collapsed attribute vectors. The VLB is defined as 
\begin{align}
L(q) = \int q(\mathbf{\Psi})\log\frac{P(\mathbf{X},\mathbf{\Psi})}{q(\mathbf{\Psi})}d\mathbf{\Psi}, 
\end{align}
where $\mathbf{\Psi}=\{\mathbf{Z}, \mathbf{\Theta}, \bm{\pi}\}$. Thus, the VLB is written as \begin{align}
L(q) &= \sum_{\mathbf{Z}}\int\int q(\mathbf{Z})q(\bm{\Theta}, \bm{\pi})\log\frac{P(\mathbf{X},\mathbf{Z}, \bm{\Theta}, \bm{\pi} \vert \mathbf{G}^{\mathrm{poly}}, \mathbf{Q}, \bm{\delta}^0, \mathbf{A}^{0}, \mathbf{B}^{0})}{q(\mathbf{Z}, \bm{\Theta}, \bm{\pi})}d\bm{\Theta}d\bm{\pi}\nonumber\\
&= \mathrm{E}_{q(\mathbf{Z})q(\bm{\Theta}, \bm{\pi})}\Bigl[\log P(\mathbf{X},\mathbf{Z}, \bm{\Theta}, \bm{\pi} \vert \mathbf{G}^{\mathrm{poly}}, \mathbf{Q}, \bm{\delta}^0, \mathbf{A}^{0}, \mathbf{B}^{0}) \Bigl] \nonumber\\
& \quad - \mathrm{E}_{q(\mathbf{Z})q(\bm{\Theta}, \bm{\pi})}\Bigl[\log q(\mathbf{Z})q(\bm{\Theta}, \bm{\pi})\Bigl] \nonumber\\
& = \mathrm{E}_{q(\mathbf{Z})q(\bm{\Theta}, \bm{\pi})}\Bigl[\log P(\mathbf{X}\vert \mathbf{Z}, \bm{\Theta}, \mathbf{G}^{\mathrm{poly}}, \mathbf{Q}) \Bigl] + \mathrm{E}_{q(\mathbf{Z})q(\bm{\Theta}, \bm{\pi})}\Bigl[\log P(\mathbf{Z} \vert \bm{\pi}) \Bigl] \nonumber\\
& \quad + \mathrm{E}_{q(\mathbf{Z})q(\bm{\Theta}, \bm{\pi})}\Bigl[\log P(\bm{\pi} \vert \bm{\delta}^0 )\Bigl] + \mathrm{E}_{q(\mathbf{Z})q(\bm{\Theta}, \bm{\pi})}\Bigl[\log P(\bm{\Theta} \vert \mathbf{A}^0 , \mathbf{B}^{0} ) \Bigl] \nonumber\\
& \quad - \mathrm{E}_{q(\mathbf{Z})}\Bigl[\log q(\mathbf{Z})\Bigl] - \mathrm{E}_{q(\bm{\Theta})}\Bigl[\log q(\bm{\Theta})\Bigl] - \mathrm{E}_{q(\bm{\pi})}\Bigl[\log q(\bm{\pi})\Bigl].
\end{align}

\clearpage
Next, we show the derivations for seven terms in Equation (31). These terms are derived as

\begin{align}
& \mathrm{E}_{q(\mathbf{Z})q(\bm{\Theta}, \bm{\pi})}\Bigl[\log P(\mathbf{X}\vert \mathbf{Z}, \bm{\Theta}, \mathbf{G}^{\mathrm{poly}}, \mathbf{Q}) \Bigl] \nonumber\\
&= \sum_{i=1}^{N}\sum_{j=1}^{J}\sum_{l^*=1}^{L^*_j}\Biggl(\sum_{l=1}^{L} g_{jl^*l}\mathrm{E}_{q(\bm{z}_i)}[z_{il}] \Biggl) \Biggl(x_{ij}\mathrm{E}_{q(\theta_{jl^*})}[\log\theta_{jl^*}] \nonumber\\
&\quad + (1-x_{ij})\mathrm{E}_{q(\theta_{jl^*})}[\log(1-\theta_{jl^*})]\Biggl) \\
& \mathrm{E}_{q(\mathbf{Z})q(\bm{\Theta}, \bm{\pi})}\Bigl[\log P(\mathbf{Z} \vert \bm{\pi}) \Bigl] = \sum_{i=1}^{N}\sum_{l=1}^{L} \mathrm{E}_{q(\bm{z}_i)}[z_{il}] \mathrm{E}_{q(\bm{\pi})}[\log\pi_{l}] \\
& \mathrm{E}_{q(\mathbf{Z})q(\bm{\Theta}, \bm{\pi})}\Bigl[\log P(\bm{\pi} \vert \bm{\delta}^0 )\Bigl] \nonumber\\
& = - \log\Biggl( \frac{\prod_{l=1}^{L}\Gamma(\delta_l^0)}{\prod_{l'=1}^{L}\Gamma(\sum_{l'=1}^{L}\delta_{l'}^0)} \Biggl) + \sum_{l=1}^{L}(\delta^0_l-1)\mathrm{E}_{q(\bm{\pi})}[\log\pi_{l}] \\
& \mathrm{E}_{q(\mathbf{Z})q(\bm{\Theta}, \bm{\pi})}\Bigl[\log P(\bm{\Theta} \vert \mathbf{A}^0 , \mathbf{B}^{0} ) \Bigl] \nonumber\\
& = -\sum_{j=1}^{J}\sum_{l^*=1}^{L^*_j}\log B(a_{jl^*}^0,b_{jl^*}^0) \nonumber\\
& \quad + \sum_{j=1}^{J}\sum_{l^*=1}^{L^*_j}\Biggl( (a_{jl^*}^0-1)\mathrm{E}_{q(\theta_{jl^*})}[\log\theta_{jl^*}] + (b_{jl^*}^0-1)\mathrm{E}_{q(\theta_{jl^*})}[\log(1-\theta_{jl^*})]\Biggl) \\
& \mathrm{E}_{q(\mathbf{Z})}\Bigl[\log q(\mathbf{Z})\Bigl] =\sum_{i=1}^{N}\sum_{l=1}^{L}\log \mathrm{E}_{q(\bm{z}_i)}[z_{il}]r_{il} \\
& \mathrm{E}_{q(\bm{\Theta})}\Bigl[\log q(\bm{\Theta})\Bigl] \nonumber\\
& = -\sum_{j=1}^{J}\sum_{l^*=1}^{L^*_j}\log B(a_{jl^*}^*,b_{jl^*}^*) \nonumber\\
& \quad + \sum_{j=1}^{J}\sum_{l^*=1}^{L^*_j}\Biggl( (a_{jl^*}^*-1)\mathrm{E}_{q(\theta_{jl^*})}[\log\theta_{jl^*}] + (b_{jl^*}^*-1)\mathrm{E}_{q(\theta_{jl^*})}[\log(1-\theta_{jl^*})]\Biggl) \\
& \mathrm{E}_{q(\bm{\pi})}\Bigl[\log q(\bm{\pi})\Bigl] = - \log\Biggl( \frac{\prod_{l=1}^{L}\Gamma(\delta_l^*)}{\prod_{l'=1}^{L}\Gamma(\sum_{l'=1}^{L}\delta_{l'}^*)} \Biggl) + \sum_{l=1}^{L}(\delta^*_l-1)\mathrm{E}_{q(\bm{\pi})}[\log\pi_{l}].
\end{align}

Then, by substituting Equations (32) to (38) with some algebra, the VLB is given as 
\begin{align}
& L(q) = \sum_{i=1}^{N}\sum_{j=1}^{J}\sum_{l^*=1}^{L^*_j}\Biggl(\sum_{l=1}^{L} g_{jl^*l}r_{il} \Biggl)\Biggl(x_{ij}\psi(a^*_{jl^*})-\psi(a^*_{jl^*}+b^*_{jl^*}) \nonumber\\
& \quad + (1-x_{ij})\psi(b^*_{jl^*})-\psi(a^*_{jl^*}+b^*_{jl^*}) \Biggl)\nonumber\\
& \quad + \sum_{i=1}^{N}\sum_{l=1}^{L}r_{il}\Biggl(\psi(\delta_l^*)-\psi(\sum_{l'=1}^{L}\delta_{l'}^*) \log r_{il} \Biggl) \nonumber\\
& \quad + \log\Biggl( \frac{\prod_{l=1}^{L}\Gamma(\delta_l^*)}{\prod_{l'=1}^{L}\Gamma(\sum_{l'=1}^{L}\delta_{l'}^*)} \Biggl)- \log\Biggl( \frac{\prod_{l=1}^{L}\Gamma(\delta_l^0)}{\prod_{l'=1}^{L}\Gamma(\sum_{l'=1}^{L}\delta_{l'}^0)} \Biggl) \nonumber\\
& \quad + \sum_{l=1}^L(\delta^0_l - \delta^*_l)\mathrm{E}_{q(\bm{\pi})}[\log\pi_{l}]\nonumber \\
& \quad + \sum_{j=1}^{J}\sum_{l^*=1}^{L^*_j}\Biggl\{\log B(a_{jl^*}^*,b_{jl^*}^*) - \log B(a_{jl^*}^0,b_{jl^*}^0) \nonumber\\
& \quad + (a_{jl^*}^0 - a_{jl^*}^*)\Bigl(\psi(a_{jl^*}^*)-\psi(a_{jl^*}^*+b_{jl^*}^*) \Bigl) \nonumber\\
& \quad + (b_{jl^*}^0-b_{jl^*}^*)\Bigl(\psi(b_{jl^*}^*)-\psi(a_{jl^*}^*+b_{jl^*}^*) \Bigl)\Biggl\},
\end{align}
where $\Gamma(\cdot)$ and $\psi(\cdot)$ denote the gamma and digamma functions, respectively.

\newpage
\subsection*{Supplementary Material B}
The true Q-matrices used in all the simulation studies were specified in the following manner. The Q-matrix with $K=4$ and $J=60$ comprises the diagonal matrices with one and two, 28 $q$-vectors that require two attributes, and 24 $q$-vectors that require three attributes. The number of items measuring each attribute is 34 for all the attributes. The Q-matrix with $K=4$ and $J=120$ is the double stacking of the Q-matrix with $K=4$ and $J=60$. Similarly, the Q-matrix with $K=7$ and $J=60$ contains the diagonal matrices with one and two, 14 $q$-vectors that require two attributes, 21 $q$-vectors that require three attributes, and 11 $q$-vectors that require four attributes. The number of items that measure each attribute was set to be approximately equal across the attributes as follows: 20 (A1), 21 (A2), 22 (A3), 23 (A4), 22 (A5), 21 (A6), and 20 (A7). The Q-matrix with $K=7$ and $J=120$ is the double stacking of the Q-matrix with $K=7$ and $J=60$.

In the following figures for the specifications of the true Q-matrices, the white, gray, and black boxes denote the Q-matrix entries that take the values of 0, 1, and 2, respectively.

\begin{figure}[htbp]
 \begin{center}
 \includegraphics[height=\textheight,keepaspectratio]{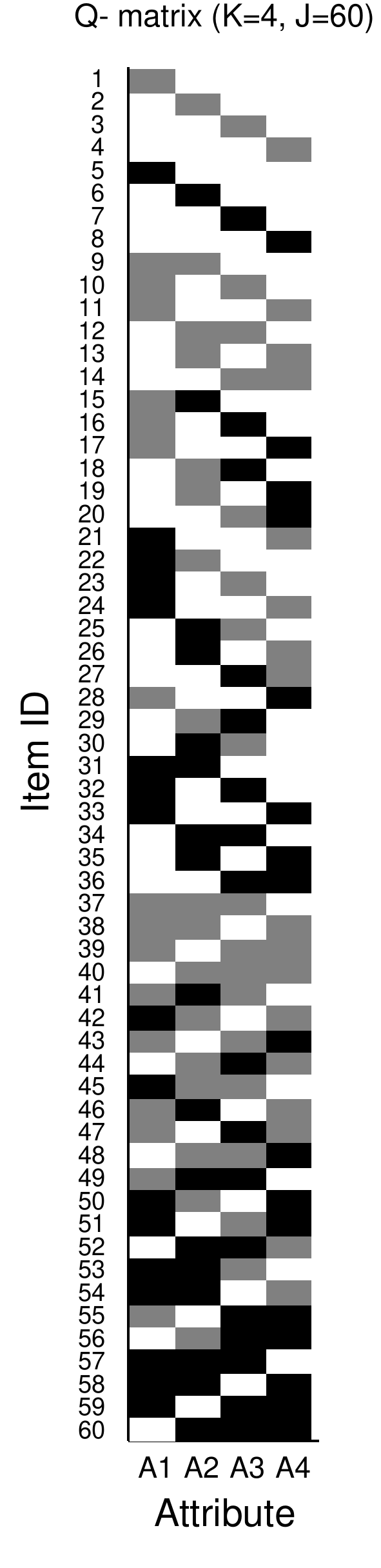}
 \end{center}
\end{figure}

\begin{figure}[htbp]
 \begin{center}
 \includegraphics[height=\textheight,keepaspectratio]{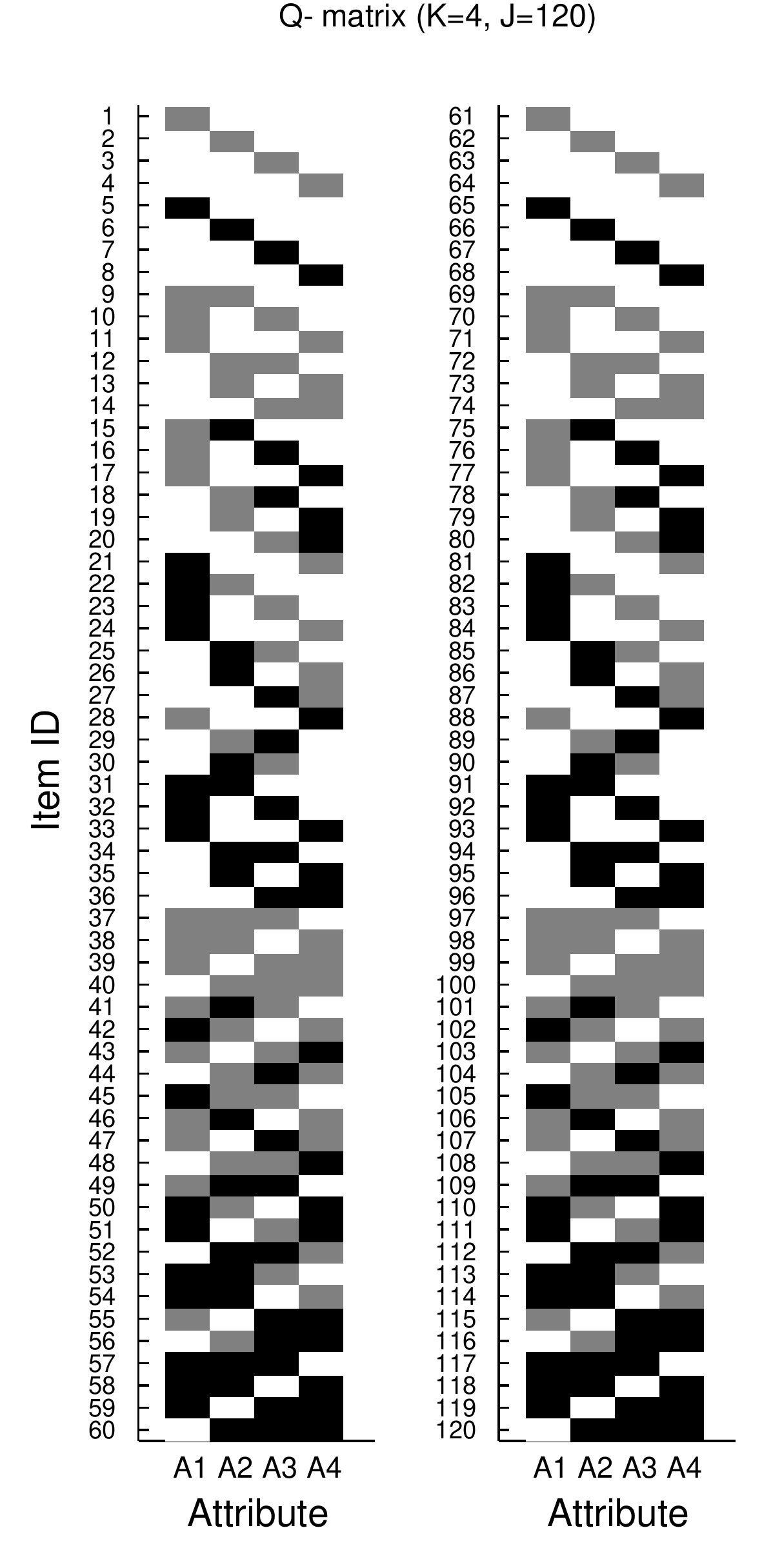}
 \end{center}
\end{figure}

\begin{figure}[htbp]
 \begin{center}
 \includegraphics[height=\textheight,keepaspectratio]{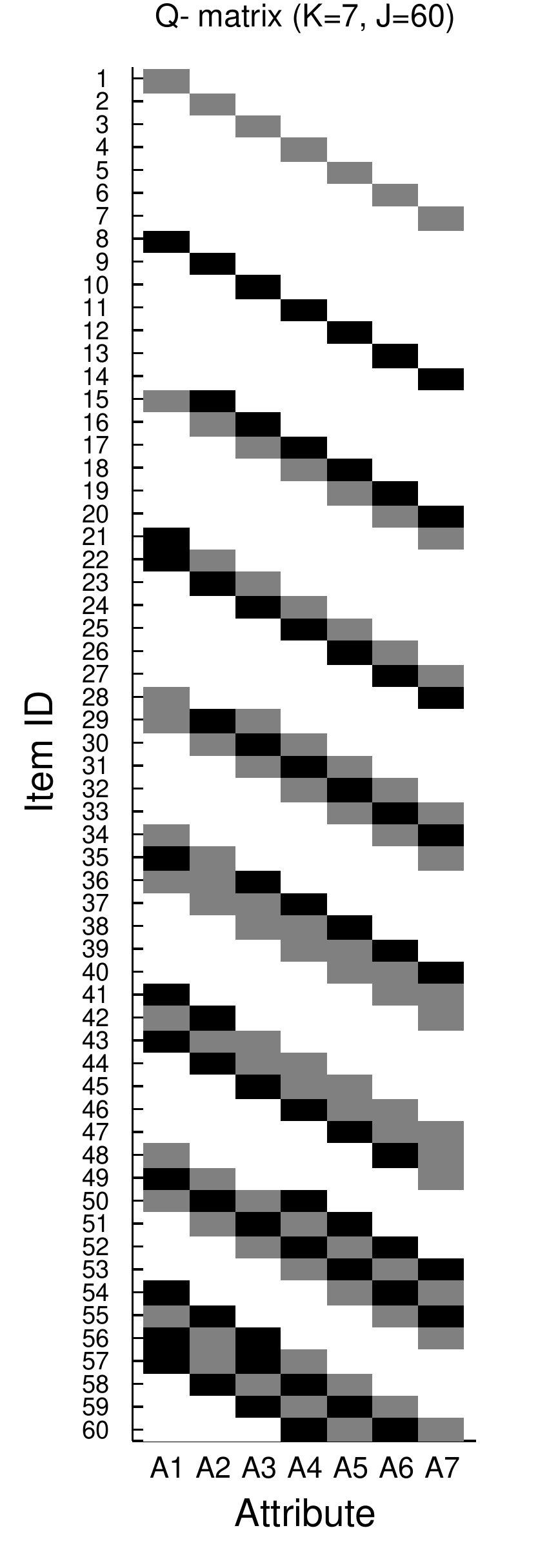}
 \end{center}
\end{figure}

\begin{figure}[htbp]
 \begin{center}
 \includegraphics[height=\textheight,keepaspectratio]{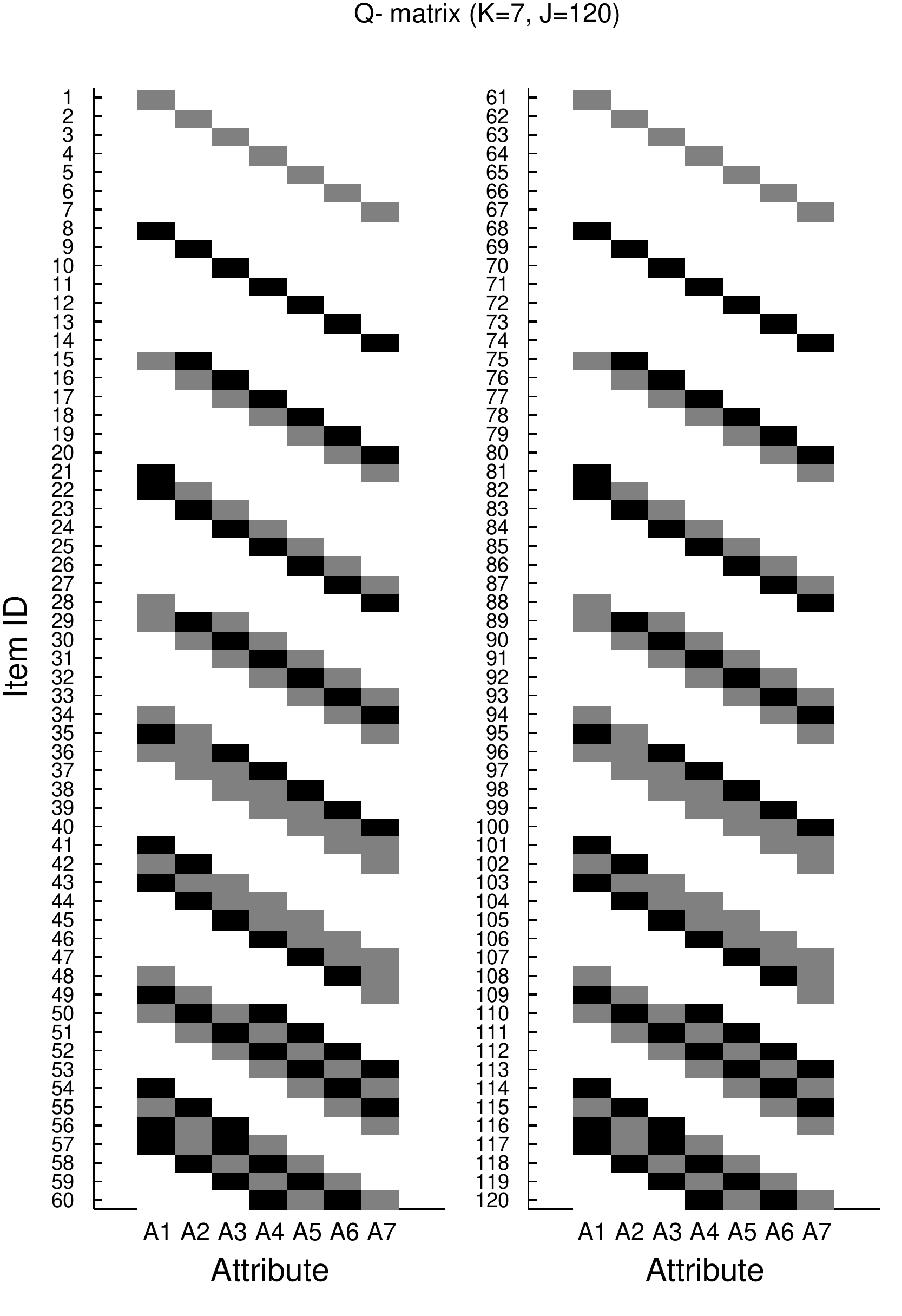}
 \end{center}
\end{figure}

\newpage
\subsection*{Supplementary Material C}
\subsection*{Additional Simulation Study: Performance of the Polytomous-Attribute Saturated DCM with its G-Matrices Using Reduced Attribute Vectors}
We investigated the performance of the polytomous-attribute saturated DCM with its G-matrix based on reduced attribute vectors. In the additional simulation study, we considered a sample size of 10000 or 30000, a number of items of 60, a correlation coefficient among the attributes of .1, and a number of attributes of $K=4$. Regarding the specification of the true values of the correct-response probability parameters, we used the same values of $p_j^{\mathrm{lowest}}$ and $p_j^{\mathrm{highest}}$ as the relevant conditions in Simulation Study 1. Then, for each item, we applied the same procedure to set the true values of these parameters given the item-specific reduced attribute vectors. The other specifications were set to be the same as those in Simulation Study 1. 

\subsubsection*{Results}
Table \ref{tab:simulation2} shows the biases and RMSEs for the correct-response probability parameters in all the conditions of the additional simulation study. Similar to Simulation Study 1, the biases and RMSEs were averaged according to the number of {attributes measured by one item}. For the polytomous-attribute saturated DCM with its G-matrices using reduced attribute vectors, the number of correct-response probability parameters for each item becomes $M^{K_j^*}$. This is in contrast to that with the G-matrices using collapsed attribute vectors, where the number of these parameters for each item is $2^{K_j^*}$.

Regarding the correct-response probability parameters, the same pattern as for collapsed attribute vectors was found. Specifically, the values of the biases were close to zero, although they were larger than those with collapsed attribute vectors. In addition, sample size lowers the values of the RMSEs and these values increased as the number of attributes measured by one item rose. In particular, the values of the RMSEs were relatively large when the sample size was 10000 and the number of attributes measured by one item was three. However, the RMSEs were less than approximately .0700 across all the conditions. With respect to the mixing proportion parameters, the absolute values of the largest positive and negative biases were within .006 and the maximum value of the RMSEs was also less than about .008. These results suggest the satisfactory accuracy of the proposed method for estimating the mixing proportion parameters.

Lastly, the performance of our method for estimating attribute mastery patterns was less satisfactory than the model with G-matrices using collapsed attribute vectors owing to more notched item-specific correct-response probabilities. Since the number of item-specific attribute mastery patterns that can have different correct-response probabilities increases to $M^{K_j^*}$ from $2^{K_j^*}$ in the model with reduced attribute vectors, the values of the correct-response probabilities among the item-specific attribute mastery patterns differ little from each other, meaning that the item's power to distinguish the likelihood at which a respondent belongs to a certain attribute mastery pattern also decreases compared with the model using collapsed attribute vectors. Hence, a large number of items would be required to ensure the accuracy of the attribute profile estimation in the model based on reduced attribute vectors.

\begin{table}[!htbp]
\
\renewcommand\thetable{C1} 
\centering
\caption{Results of the additional simulation study ($K=4$, correlation coefficient .1)}
\resizebox{\linewidth}{!}{
\begin{tabular}{cccccccccc}
\toprule
\multicolumn{10}{c}{Biases and RMSEs for the correct-response probability parameters}\\
\toprule
& & \multicolumn{8}{c}{Number of Attributes Measured by One Item} \\
\cmidrule{3-10}& & \multicolumn{2}{c}{1} & & \multicolumn{2}{c}{2} & & \multicolumn{2}{c}{3} \\
\cmidrule{1-4}\cmidrule{6-7}\cmidrule{9-10}Number of Items & Sample Size & Bias& RMSE& & Bias& RMSE& & Bias& RMSE \\
\midrule
\multirow{2}[2]{*}{60} & 10000 & -.0021 & .0228 & & -.0021 & .0327 & & -.0019 & .0701 \\
& 30000 & -.0012 & .0093 & & -.0010 & .0149 & & -.0010 & .0325 \\
\bottomrule \\
\\
\end{tabular}%
}

\resizebox{\linewidth}{!}{
\begin{tabular}{ccccccc}
\multicolumn{7}{c}{Maximum and minimum values of the biases and RMSEs for the mixing proportion parameters}\\ \toprule
& & \multicolumn{5}{c}{Mixing Proportion Parameter} \\
\cmidrule{3-7}& & \multicolumn{2}{c}{Bias} & & \multicolumn{2}{c}{RMSE} \\
\cmidrule{3-4}\cmidrule{6-7}Number of Items & Sample Size & Maximum & Minimum & & Maximum & Minimum \\
\midrule
\multirow{2}[2]{*}{60} & 10000 & .0032 & -.0059 & & .0081 & .0027 \\
& 30000 & .0008 & -.0010 & & .0035 & .0012 \\
\bottomrule \\
\\
\end{tabular}%
}

\resizebox{\linewidth}{!}{
\begin{tabular}{ccccccc}
\multicolumn{7}{c}{EACRs and PACRs for the attribute profile estimation}\\
\toprule 
& & \multicolumn{5}{c}{Attribute Classification Rate} \\
\cmidrule{3-7}Number of Items & Sample Size & $\text{EACR}_{\alpha_1}$ & $\text{EACR}_{\alpha_2}$ & $\text{EACR}_{\alpha_3}$ & $\text{EACR}_{\alpha_4}$ & $\text{PACR}_{\bm{\alpha}}$ \\
\midrule
\multirow{2}[2]{*}{60} & 10000 & .757& .761& .760& .776& .370 \\
& 30000 & .783& .786& .784& .795& .419 \\
\bottomrule
\end{tabular}%
}
 \begin{tablenotes}
 \item \footnotesize{\textit{Note.} Maximum and minimum values of the biases correspond to the largest positive and negative biases among the $M^K$ mixing proportion parameters, respectively. RMSE = root-mean-square error, EACR = element-wise attribute classification rate, PACR = pattern-wise attribute classification rate.}
 \end{tablenotes}
\label{tab:simulation2}%
\end{table}%

\clearpage
\subsection*{Supplementary Material D}
\subsection*{Empirical Study: Estimates of the Correct-response Probability Parameters}
Tables \ref{tab:realdataGibbs} and \ref{tab:realdataVB} present the estimates of the EAP and posterior standard deviation (SD) for the correct-response probability parameters from the Gibbs sampler and the proposed method, respectively. In the tables, $P(\cdot)$ means the correct-response probability given collapsed attribute vectors for the items. For instance, when item $j$ has $q_j=(1,0,0)$, its collapsed attribute vectors are $\alpha_{j1}^{**} = (0)$ and $\alpha_{j2}^{**} = (1)$. Hence, $P(0)$ for item $j$ corresponds to $P(x_{ij}=1|\alpha_{j1}^{**} = (0))$.

\begin{table}[htbp]
\renewcommand\thetable{D1} 
\centering
\caption{The estimates of EAP and posterior standard deviation for the correct-response probability parameters from the Gibbs sampler}
\resizebox{\linewidth}{!}{
\begin{tabular}{ccccccccccccccccc}
\toprule
& \multicolumn{16}{c}{Gibbs Sampler} \\
\cmidrule{2-17}& \multicolumn{2}{c}{P(0)} & \multicolumn{2}{c}{P(1)} & \multicolumn{2}{c}{} & \multicolumn{2}{c}{} & & & & & & & &\\
& \multicolumn{2}{c}{P(00)} & \multicolumn{2}{c}{P(01)} & \multicolumn{2}{c}{P(10)} & \multicolumn{2}{c}{P(11)} & & & & & & & &\\
\multicolumn{1}{l}{Items} & \multicolumn{2}{c}{P(000)} & \multicolumn{2}{c}{P(001)} & \multicolumn{2}{c}{P(010)} & \multicolumn{2}{c}{P(011)} & \multicolumn{2}{c}{P(100)} & \multicolumn{2}{c}{P(101)} & \multicolumn{2}{c}{P(110)} & \multicolumn{2}{c}{P(111)} \\
\midrule
1 & .358& (.007) & .931& (.002) & & & & & & & & & & & &\\
2 & .479& (.007) & .935& (.002) & & & & & & & & & & & &\\
3 & .285& (.006) & .852& (.003) & & & & & & & & & & & &\\
4 & .457& (.007) & .946& (.002) & & & & & & & & & & & &\\
5 & .360& (.007) & .931& (.002) & & & & & & & & & & & &\\
6 & .051& (.003) & .758& (.004) & & & & & & & & & & & &\\
7 & .121& (.005) & .818& (.003) & & & & & & & & & & & &\\
8 & .227& (.006) & .603& (.008) & .506& (.025) & .879& (.004) & & & & & & & &\\
9 & .310& (.004) & .825& (.004) & & & & & & & & & & & &\\
10& .199& (.006) & .820& (.007) & .801& (.020) & .963& (.002) & & & & & & & &\\
11& .054& (.003) & .161& (.022) & .213& (.006) & .750& (.005) & & & & & & & &\\
12& .593& (.005) & .978& (.002) & & & & & & & & & & & &\\
13& .479& (.005) & .909& (.003) & & & & & & & & & & & &\\
14& .210& (.006) & .159& (.060) & .433& (.033) & .715& (.129) & .344& (.019) & .597& (.009) & .790& (.030) & .933& (.003) \\
15& .364& (.007) & .448& (.070) & .644& (.031) & .768& (.118) & .425& (.019) & .656& (.008) & .841& (.027) & .929& (.003) \\
16& .374& (.007) & .523& (.069) & .642& (.032) & .530& (.141) & .478& (.019) & .656& (.008) & .775& (.031) & .918& (.003) \\
17& .096& (.004) & .661& (.030) & .591& (.007) & .909& (.003) & & & & & & & &\\
18& .272& (.006) & .467& (.029) & .529& (.007) & .859& (.004) & & & & & & & &\\
19& .007& (.001) & .108& (.018) & .301& (.007) & .798& (.005) & & & & & & & &\\
20& .082& (.004) & .180& (.022) & .124& (.005) & .499& (.006) & & & & & & & &\\
21& .059& (.003) & .483& (.030) & .358& (.007) & .921& (.003) & & & & & & & &\\
22& .402& (.007) & .863& (.020) & .807& (.005) & .946& (.003) & & & & & & & &\\
23& .238& (.006) & .634& (.029) & .478& (.007) & .854& (.004) & & & & & & & &\\
24& .245& (.006) & .783& (.004) & & & & & & & & & & & &\\
25& .387& (.004) & .657& (.006) & & & & & & & & & & & &\\
26& .504& (.006) & .806& (.004) & & & & & & & & & & & &\\
27& .617& (.007) & .941& (.014) & .870& (.004) & .968& (.002) & & & & & & & &\\
28& .256& (.006) & .816& (.066) & .576& (.033) & .539& (.147) & .346& (.019) & .635& (.008) & .527& (.040) & .867& (.004) \\
29& .198& (.006) & .540& (.072) & .461& (.032) & .503& (.142) & .246& (.018) & .552& (.009) & .439& (.041) & .873& (.004) \\
30& .532& (.007) & .965& (.024) & .919& (.019) & .885& (.088) & .664& (.018) & .865& (.006) & .784& (.030) & .980& (.002) \\
31& .009& (.001) & .115& (.005) & .049& (.012) & .558& (.006) & & & & & & & &\\
32& .151& (.003) & .329& (.005) & & & & & & & & & & & &\\
33& .024& (.002) & .466& (.007) & & & & & & & & & & & &\\
34& .355& (.004) & .779& (.006) & & & & & & & & & & & &\\
\bottomrule
\end{tabular}%
}
 \begin{tablenotes}
 \item \footnotesize{\textit{Note.} We present the values of EAP and posterior standard deviation as ``EAP (Posterior SD).''}
 \end{tablenotes}
\label{tab:realdataGibbs}
\end{table}%

\begin{table}[htbp]
\renewcommand\thetable{D2} 
\centering
\caption{The estimates of EAP and posterior standard deviation for the correct-response probability parameters from the proposed method}
\resizebox{\linewidth}{!}{
\begin{tabular}{ccccccccccccccccc}
\toprule
& \multicolumn{16}{c}{Variational Bayes} \\
\cmidrule{2-17}& \multicolumn{2}{c}{P(0)} & \multicolumn{2}{c}{P(1)} & \multicolumn{2}{c}{} & \multicolumn{2}{c}{} & & & & & & & &\\
& \multicolumn{2}{c}{P(00)} & \multicolumn{2}{c}{P(01)} & \multicolumn{2}{c}{P(10)} & \multicolumn{2}{c}{P(11)} & & & & & & & &\\
\multicolumn{1}{l}{Items} & \multicolumn{2}{c}{P(000)} & \multicolumn{2}{c}{P(001)} & \multicolumn{2}{c}{P(010)} & \multicolumn{2}{c}{P(011)} & \multicolumn{2}{c}{P(100)} & \multicolumn{2}{c}{P(101)} & \multicolumn{2}{c}{P(110)} & \multicolumn{2}{c}{P(111)} \\
\midrule
1 & .358& (.006) & .931& (.002) & & & & & & & & & & & &\\
2 & .478& (.006) & .935& (.002) & & & & & & & & & & & &\\
3 & .285& (.006) & .852& (.003) & & & & & & & & & & & &\\
4 & .456& (.006) & .946& (.002) & & & & & & & & & & & &\\
5 & .359& (.006) & .931& (.002) & & & & & & & & & & & &\\
6 & .051& (.003) & .758& (.003) & & & & & & & & & & & &\\
7 & .121& (.004) & .818& (.003) & & & & & & & & & & & &\\
8 & .226& (.005) & .602& (.007) & .506& (.016) & .879& (.004) & & & & & & & &\\
9 & .310& (.004) & .825& (.004) & & & & & & & & & & & &\\
10& .198& (.005) & .819& (.005) & .802& (.013) & .963& (.002) & & & & & & & &\\
11& .054& (.003) & .162& (.016) & .213& (.005) & .750& (.005) & & & & & & & &\\
12& .593& (.004) & .978& (.002) & & & & & & & & & & & &\\
13& .479& (.004) & .909& (.003) & & & & & & & & & & & &\\
14& .210& (.005) & .159& (.028) & .432& (.023) & .719& (.066) & .343& (.013) & .597& (.007) & .789& (.018) & .933& (.003) \\
15& .364& (.006) & .450& (.039) & .644& (.022) & .781& (.061) & .424& (.013) & .656& (.007) & .841& (.017) & .929& (.003) \\
16& .374& (.006) & .523& (.039) & .643& (.022) & .520& (.074) & .478& (.013) & .656& (.007) & .774& (.019) & .918& (.003) \\
17& .095& (.004) & .662& (.021) & .591& (.006) & .909& (.003) & & & & & & & &\\
18& .272& (.006) & .467& (.022) & .528& (.006) & .859& (.004) & & & & & & & &\\
19& .007& (.001) & .108& (.014) & .301& (.006) & .797& (.004) & & & & & & & &\\
20& .082& (.004) & .180& (.017) & .124& (.004) & .499& (.005) & & & & & & & &\\
21& .058& (.003) & .484& (.022) & .358& (.006) & .921& (.003) & & & & & & & &\\
22& .402& (.006) & .862& (.015) & .807& (.005) & .946& (.002) & & & & & & & &\\
23& .238& (.006) & .635& (.021) & .478& (.006) & .854& (.004) & & & & & & & &\\
24& .245& (.005) & .783& (.004) & & & & & & & & & & & &\\
25& .387& (.004) & .657& (.005) & & & & & & & & & & & &\\
26& .503& (.006) & .806& (.003) & & & & & & & & & & & &\\
27& .617& (.006) & .941& (.010) & .870& (.004) & .968& (.002) & & & & & & & &\\
28& .256& (.006) & .816& (.030) & .575& (.022) & .544& (.074) & .345& (.013) & .635& (.007) & .525& (.023) & .866& (.004) \\
29& .198& (.005) & .537& (.039) & .462& (.023) & .492& (.074) & .246& (.011) & .552& (.007) & .438& (.022) & .873& (.004) \\
30& .532& (.007) & .970& (.013) & .917& (.013) & .903& (.044) & .663& (.013) & .865& (.005) & .783& (.019) & .980& (.002) \\
31& .009& (.001) & .115& (.004) & .047& (.007) & .558& (.005) & & & & & & & &\\
32& .151& (.003) & .329& (.005) & & & & & & & & & & & &\\
33& .024& (.001) & .466& (.005) & & & & & & & & & & & &\\
34& .355& (.004) & .779& (.005) & & & & & & & & & & & &\\
\bottomrule
\end{tabular}%
}
 \begin{tablenotes}
 \item \footnotesize{\textit{Note.} We present the values of EAP and posterior standard deviation as ``EAP (Posterior SD).''}
 \end{tablenotes}
\label{tab:realdataVB}
\end{table}%

\end{document}